\documentclass[fleqn,10pt]{wlscirep}
\usepackage[utf8]{inputenc}
\usepackage[T1]{fontenc}
\usepackage{subcaption}
\usepackage{float}
\usepackage{booktabs}
\usepackage{multirow}
\usepackage{graphicx}
\usepackage{adjustbox}
\usepackage{CJKutf8}

\title{A Multi-Scale Spatial Attention-Based Zero-Shot Learning Framework for Low-Light Image Enhancement}

\author[1,*]{Muhammad Azeem Aslam}
\author[2]{Hassan Khalid}
\author[3]{Nisar Ahmed}
\affil[1]{School of Information Engineering, Xi'an Eurasia University, Xi'an,  710065, Shaanxi, China.}
\affil[2]{Department of Electrical Engineering, University of Engineering and Technology Lahore, Lahore, 54890, Punjab, Pakistan.}
\affil[3]{Department of Computer Science, University of Engineering and Technology Lahore, (New Campus), Lahore, 54890, Punjab, Pakistan.}

\affil[*]{Email: $engr\_hassan09@ymail.com$}

\begin{abstract}
Low-light image enhancement remains a challenging task, particularly in the absence of paired training data. In this study, we present LucentVisionNet, a novel zero-shot learning framework that addresses the limitations of traditional and deep learning-based enhancement methods. The proposed approach integrates multi-scale spatial attention with a deep curve estimation network, enabling fine-grained enhancement while preserving semantic and perceptual fidelity. To further improve generalization, we adopt a recurrent enhancement strategy and optimize the model using a composite loss function comprising six tailored components, including a novel no-reference image quality loss inspired by human visual perception. Extensive experiments on both paired and unpaired benchmark datasets demonstrate that LucentVisionNet consistently outperforms state-of-the-art supervised, unsupervised, and zero-shot methods across multiple full-reference and no-reference image quality metrics. Our framework achieves high visual quality, structural consistency, and computational efficiency, making it well-suited for deployment in real-world applications such as mobile photography, surveillance, and autonomous navigation.
\end{abstract}
\begin{document}

\flushbottom
\maketitle
%
%
\thispagestyle{empty}

\section{Introduction}
\label{sec:int}
\begin{figure}[htbp]
    \centering
    \begin{subfigure}[b]{0.19\textwidth}
        \includegraphics[width=\linewidth]{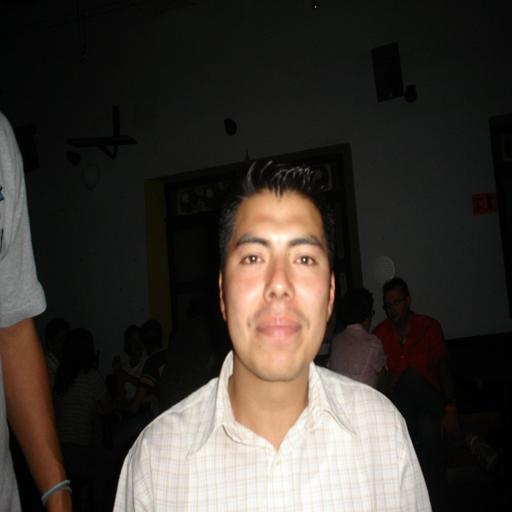}
        \caption{Input}
    \end{subfigure}
    \begin{subfigure}[b]{0.19\textwidth}
        \includegraphics[width=\linewidth]{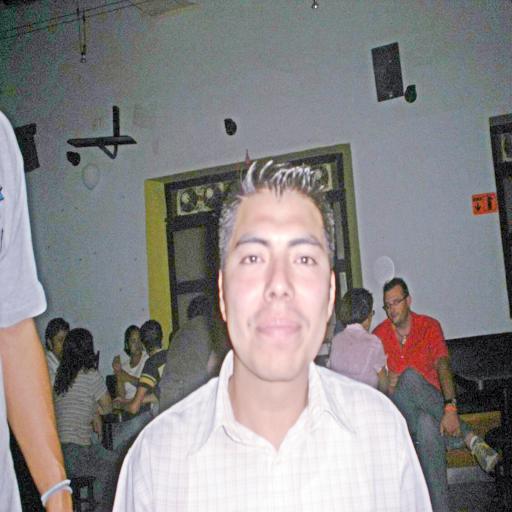}
        \caption{Zero DCE}
    \end{subfigure}
    \begin{subfigure}[b]{0.19\textwidth}
        \includegraphics[width=\linewidth]{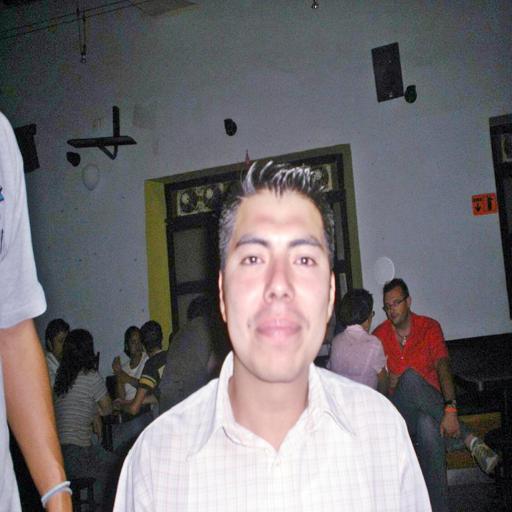}
        \caption{Zero DCE++}
    \end{subfigure}
    \begin{subfigure}[b]{0.19\textwidth}
        \includegraphics[width=\linewidth]{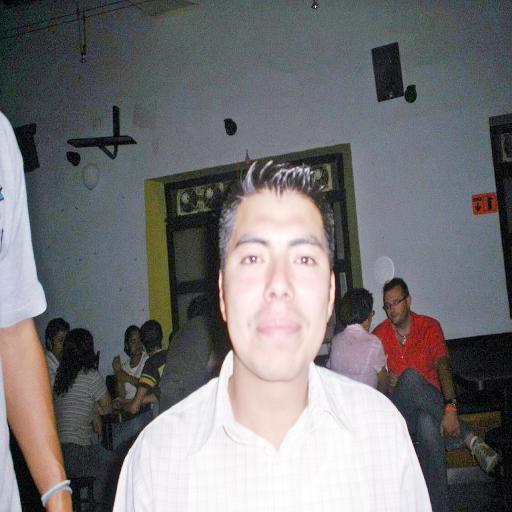}
        \caption{Semantic Guided }
    \end{subfigure}
    \begin{subfigure}[b]{0.19\textwidth}
        \includegraphics[width=\linewidth]{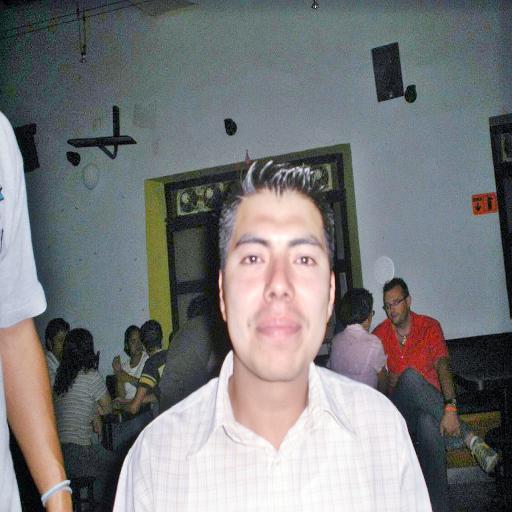}
        \caption{Ours}
    \end{subfigure}
    \caption{A visual comparison of enhanced images generated by various zero-shot learning algorithms and the proposed \emph{LucentVisionNet} model reveals that the proposed approach demonstrates superior performance. Specifically, \emph{LucentVisionNet} achieves more adaptive enhancement in terms of brightness, contrast, and perceptual quality, thereby outperforming existing methods.}
    \label{fig:five-images}
\end{figure}

\begin{figure}[htb]
    \centering
    \includegraphics[width=\linewidth]{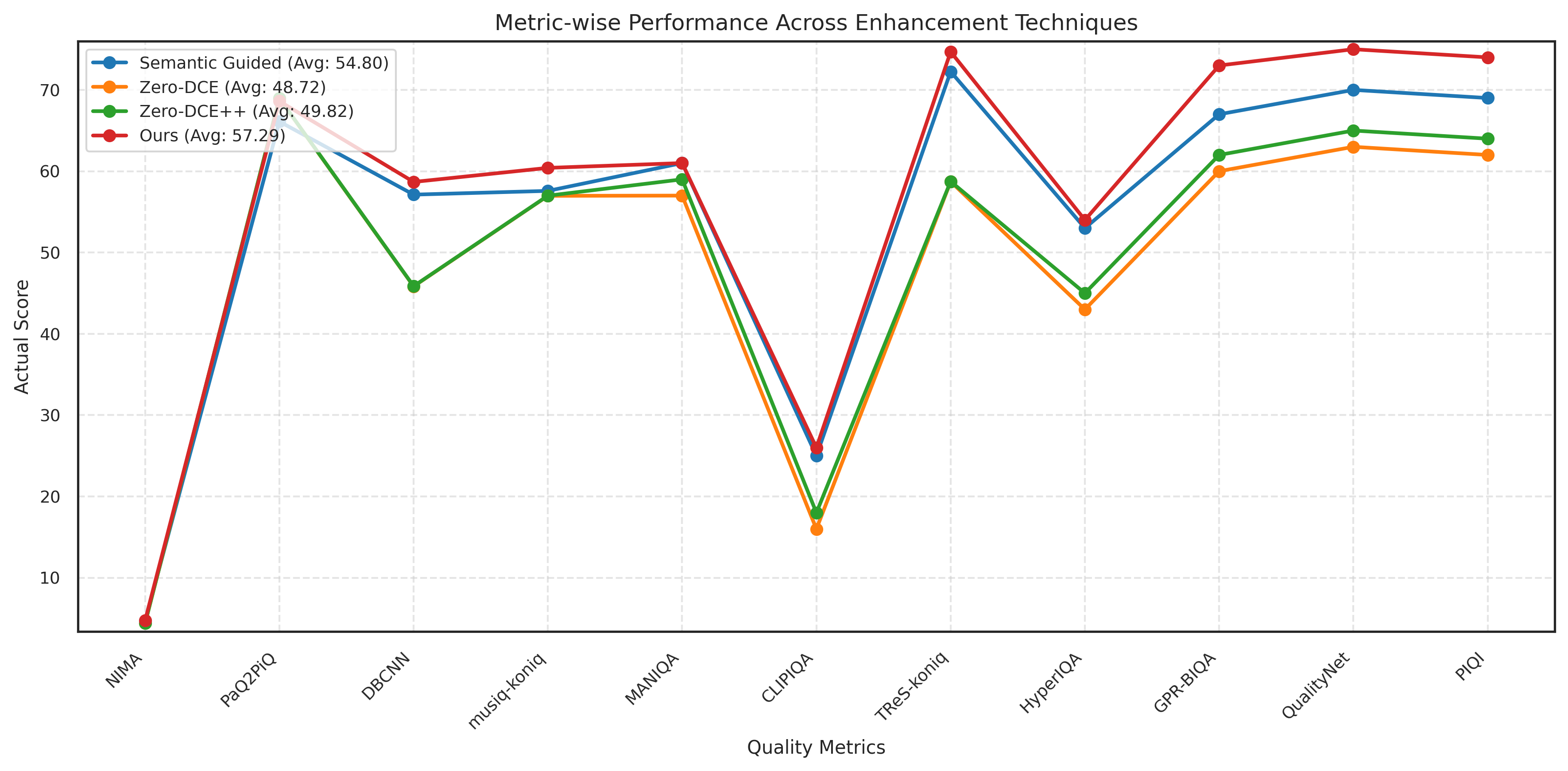}
    \caption{Perceptual Objective Evaluation score for the same image generated by different ZSL algorithms}
    \label{MOS}
\end{figure}

Image acquisition is not always carried out under ideal conditions in terms of camera characteristics, ambient conditions, or acquisition angle. Poor illumination is one of the most prevalent and limiting problems for digital images \cite{[1],[2]}, particularly in indoor acquisition environments, at night, or due to camera constraints. The outcome is dark, noisy, low contrast, and poor perceptual quality images. This affects both the human perceptual experience and the performance of high-level semantic tasks such as object recognition, segmentation, and depth estimation. Possible solutions include modifying the surrounding environment or making changes to the camera characteristics, such as increasing the ISO or prolonging the exposure time. Ambient conditions can be improved only in controlled environments and are therefore not generally applicable. Moreover, increasing the ISO leads to higher noise levels—primarily due to read noise, thermal noise, shot noise, and other contributing factors. Similarly, increasing the exposure time might result in increased thermal noise, motion blur, camera shake, and overexposure, making it an unsuitable solution as well. Image editing software can also be used to enhance low-light images \cite{[3]}; however, it has two key drawbacks. First, using such software requires expertise and can be time-consuming. Second, this approach often lacks automation, consistency, and speed, necessitating manual fine-tuning.\\
Quality of images is of vital importance for human visual experience \cite{ahmed2019image, aslam2023vrl} as well as many high-level computer vision tasks such as autonomous driving, surveillance, scientific or medical imaging where preservation of semantic information is vital for interpretation and decision making \cite{[4],[5],[6],[7],[8]}. For instance, low-light images may make it difficult to perform object recognition, anomaly identification, or face identification during surveillance or affect the visibility of the road and its surroundings, which is critical for safe navigation \cite{[9], [10], aslam2024tqp}.\\
Automated low-light image enhancement techniques can rescue the situation by improving the perceptual and semantic quality of these images. This leads to an enhanced perceptual experience, improved extraction of semantic information, and greater accuracy in interpretation \cite{[5], [11], [12]}. Among the most striking benefits of automated low-light image enhancement methods are speed, consistency, and scalability. The problem is addressed via both traditional image processing algorithms and modern deep learning-based solutions. The approaches work by preserving the content and reducing artifacts, with an ability to be integrated into existing systems and making them compatible with a wide range of applications.\\
Among the traditional techniques, histogram-based methods \cite{[13],[14],[15],[16]}, exposure correction, image fusion, and Retinex-based methods are most prominent. Gamma correction \cite{[17]} and tone mapping \cite{[18], [19]} are the most common methods to perform exposure correction. Histogram equalization and its variations, including BPDHE (Brightness Preserving Dynamic Histogram Equalization) \cite{[16]} and CLACHE (Contrast Limited Adaptive Histogram Equalization) \cite{[15]}. Combining multiple images of a scene taken under various exposure conditions is known as image fusion \cite{[20], [21]}. This can be done via weighted averaging, wavelet fusion \cite{[22]}, or Laplacian pyramid fusion \cite{[23]}. Retinex is another well-known non-linear technique \cite{[24]} for improving images in low light.  By first breaking down the image into its illumination and reflectance components, then improving the illumination component, the technique makes features in low-light images visibly better. The most effective non-deep learning technique for enhancing the low-light qualities of digital images is multi-scale Retinex \cite{[25]}, adaptive Retinex \cite{[26]}, color Retinex \cite{[27]}, and multi-scale Retinex with color restoration \cite{[28]}.\\
In recent years, deep learning-based methods for improving low-light images have become more and more popular \cite{[29]}. They outperform conventional methods in terms of perceived quality.  Their exceptional performance can be attributed to their capacity to extract intricate features from large amounts of training data.\\
Unsupervised techniques, as opposed to supervised ones, might learn straight from the input images without the need for any ground truth information.  Generative adversarial networks (GANs) are commonly employed to enhance low-light images [30]. GANs consist of two networks: a discriminator network that can distinguish between produced and genuine images, and a generator network that enhances preexisting images. The discriminator network is fooled by high-quality augmented images generated by the generator network, which has been trained for this purpose. Supervised methods \cite{[31],[32],[33],[34]} perform learning from image-to-image mapping and have provided the highest scores in terms of quality metrics on benchmark datasets \cite{[35],[36],[37],[38],[39],[40],[41],[42]}. The limitation for these approaches lies with their dependence on paired training examples with low-light or perfectly lit images of the same scene. Acquisition or preparation of a dataset with such image pairs is expensive or sometimes infeasible due to weaker control over ambient conditions. Moreover, supervised methods trained on a particular type of lightning conditions can perform image enhancement in similar scenario only.\\
On the other hand, unsupervised methods may involve hyperparameter optimization \cite{[42],[43],[44]}, Retinex-based learning approaches \cite{[42],[43],[44],[45]}, or zero-shot learning techniques \cite{[46]}. These methods do not require paired training samples and rely solely on low-light images. These approaches have their flaws, such as serious noise amplification, poor adaptive enhancement, and a large number of model parameters. While image brightening enhances visibility, it also amplifies noise substantially; applying denoising techniques afterward may degrade or remove critical semantic details. Poor adaptivity results in overexposed areas in the image, which result in brightening the already bright areas and therefore provide unsatisfactory perceptual quality. An excessive number of model parameters presents a significant challenge to adoption, as some of the most successful models are too complex for deployment in many real-world scenarios, ultimately restricting their applicability.\\
Zero-shot learning (ZSL) has recently emerged as a promising direction in the field of low-light image enhancement, enabling image correction without the need for paired training data. One of the pioneering efforts in this domain is Zero-DCE \cite{[47]}, which was further improved upon by Zero-DCE++ \cite{Zero-DCE++}, and later extended through semantic-guided zero-shot learning approaches \cite{ZeroDCEsemantic}. While the semantic-guided ZSL approach integrates semantic cues to improve quality of enhanced images, it still exhibits key limitations such as a lack of perceptual learning aligned with human visual preferences, insufficient attention mechanisms for curve estimation, and limited ability to capture both fine and coarse details due to the absence of multi-scale learning strategies.\\
This study introduces a novel ZSL-based framework for low-light image enhancement that systematically addresses the limitations of existing approaches. The proposed method integrates multi-scale curve estimation with spatial attention mechanisms and incorporates both perceptual and semantic guidance, thereby enabling the generation of visually compelling and perceptually coherent enhancements. Notably, the framework operates in an unsupervised manner, eliminating the need for paired training data, and exhibits strong generalization across a wide range of illumination conditions. To further enhance the model’s generalization capability, a recurrent image enhancement strategy is employed and optimized using six loss functions. Five of these are adopted from existing ZSL–based methodologies \cite{[47], Zero-DCE++, ZeroDCEsemantic}, while the sixth is a novel loss function proposed in this work \cite{musiq}, based on a no-reference image quality assessment (NR-IQA) algorithm. This additional loss term is specifically designed to improve perceptual fidelity and enhance the preservation of fine-grained structural details, aligning with human visual perception, particularly under extreme low-light conditions. Extensive experimental evaluations—both qualitative and quantitative—are conducted to benchmark the proposed framework against state-of-the-art techniques. The empirical results substantiate the superiority of our model in terms of visual quality, semantic consistency, and computational efficiency. In particular, the model demonstrates real-time capability, processing images of $1200\times900$ resolution in approximately 1 to 1.5 seconds on a single GPU, underscoring its practical applicability for both still image and video enhancement tasks in low-light environments.\\
Figure \ref{fig:five-images} presents a visual comparison of low-light enhanced example images produced by all ZSL algorithms under consideration. Notably, the proposed \emph{LucentVisionNet} demonstrates superior visual quality in comparison to existing methods. In addition, Figure \ref{MOS} reports the average score calculated by averaging scores for all Blind Image Quality assessment metrics (score scaled to 100) for the same set of images. The proposed algorithm achieves the highest score, indicating its effectiveness in producing perceptually favorable results.

\section{Related Work}
Exposure correction, histogram equalization, image fusion, dehazing, and Retinex-based techniques are common practices to improve low-light images. These methods can improve the contrast and perceptual appearance of images, but may result in increased noise or low color restoration. Moreover, these approaches are not learning-based and perform sub-optimally for several high-level computer vision tasks.\\
Learning-based solutions are mostly based on deep learning algorithms and provide superior low-light image enhancement performance in terms of perceptual appearance and quality metrics. These methods can be broadly classified into supervised \cite{[31],[32],[33],[34]} and unsupervised methods \cite{[47],[48],[49],[50]}. Supervised learning-based solutions provide the highest performance in terms of quality metrics on the benchmark datasets \cite{[35],[36],[37],[38],[39],[40],[41],[42]} as compared to unsupervised approaches. However, in contrast to other supervised learning tasks, they are trained using paired images that are not consistent (absolute) for a task. For instance, a low-light scene can have multiple high-light variants, making it difficult to determine the most optimal reference image. The selection of the ideal reference image \cite{[51], [52]} remains a challenge after correction/selection by experts, therefore increasing the complexity of the problem and the reliability of a solution.\\ 
Therefore, the most prominent challenge in supervised low-light enhancement methods is the presence of multiple potential references. A solution to these problems is the MAXIM \cite{[53]} which is a large and complex network and has state-of-the-art performance. The problem with such methods is the computational complexity, which makes them time-consuming and may limit their applicability to some scenarios. Another type of supervised approach is the use of hyperparameters \cite{[42],[43],[44]} or Retinex \cite{[42],[43],[44], [45]} during training to connect the input image to the output.\\
As an example of a hyperparameter-based approach, Fu et al. \cite{[44]} proposed the use of a sub-network to perform automatic selection of hyperparameters, whereas Chen et al. \cite{[43]} introduced the use of the exposure time ratio between the reference and low-light image as hyperparameters. Going towards Retinex-based supervised learning, Wei et al. \cite{[45]} implemented the streamlined version of the Retinex model into the network. The streamlined Retinex model assumes that all three-color channels share the same illumination image, however, this assumption is at contrast with reality \cite{[51]}, resulting in unsatisfactory denoising results. To overcome these limitations, Zhang et al. \cite{[42], [52]} presented a hybrid approach and incorporated both Retinex and hyperparameters into their network to perform color correction and noise removal in the reflection image. Despite their relatively less computational complexity, they are still slow and may not be suitable for some real-time application requirements.\\
Unsupervised methods are based on the assumption that the output image satisfies certain constraints and therefore makes them stable for unseen scenarios. For instance, Guo et al. \cite{[47]} proposed a specifically designed loss function based on the constraint of having a mean brightness between 0.4 to 0.6. This mean value assumption makes it incredibly simple and fast, but makes it unsuitable to restore color information or remove noise. Xiong et al. \cite{[54]} make a constraint on the initial value of the illumination image in a simplified Retinex model. Their constraint is based on the assumption that the maximum value in each of the red, green and blue channle is the initial value of the illumination image. Jiang et al. \cite{[48]} make use of GAN model to learn a constraint on the output from normal light images. Similarly, Ma et al. \cite{[49]} constraints the similarity of the outputs throughout the training process. These models meet the complexity requirements for most applications but lack in producing visually appealing and perceptually accurate results.\\
ZSL algorithms have emerged as a transformative approach for low-light image enhancement, enabling models to improve brightness, contrast, and color fidelity without paired training data. Techniques like Zero-Reference Deep Curve Estimation (ZRDCE) utilize deep neural networks to predict pixel-wise adjustment curves, dynamically enhancing images without reference to ground-truth normal-light images, making it ideal for real-time applications such as mobile photography \cite{[47]}. Similarly, Semantic-Guided Zero-Shot Learning (SG-ZSL) integrates semantic information, such as object categories and scene context, to guide enhancement, preserving meaningful content and achieving superior perceptual quality in complex scenes like autonomous driving footage \cite{ZeroDCEsemantic}. These ZSL methods demonstrate robust generalization to unseen lighting conditions, outperforming traditional supervised approaches in flexibility and practicality \cite{[46]}.\\
While recent deep learning-based solutions have provided good performance in various low-light enhancement scenarios, they sometimes suffer from noise amplification and overexposed scenarios in extremely dark and bright regions of the image. Moreover, the computational complexity of these methods is another limiting concern, as large model sizes limit the applicability of these models in many scenarios. This work proposes a multiscale model with spatial attention and perceptual as well as semantic guidance to produce aesthetically appealing and perceptually accurate results with low computational cost to overcome these limitations.
\section{Proposed Model}
\begin{figure}[htbp]
    \centering
    \begin{subcaptionbox}{Multi-Scale Spatial Curve Estimation Network\label{fig:img1arch}}{
        \includegraphics[width=0.9\linewidth]{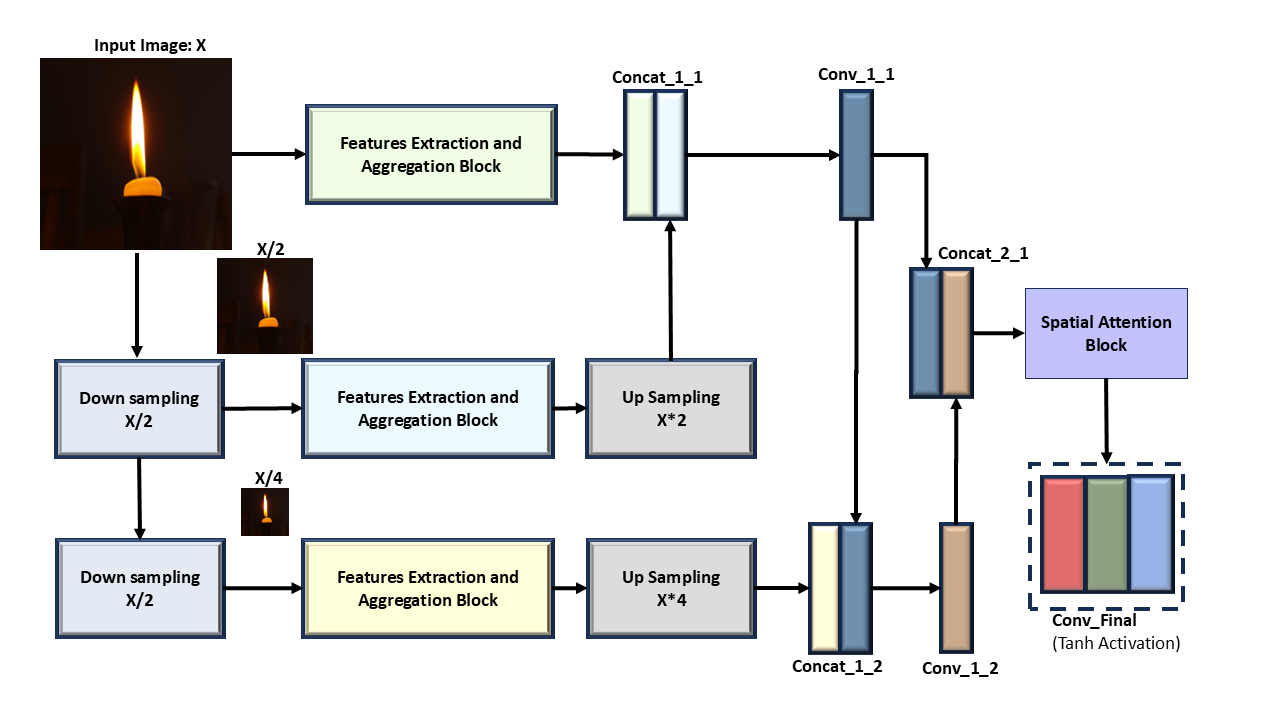}
    }\end{subcaptionbox}

    \vspace{0.3cm}

    \begin{subcaptionbox}{Feature Extraction and Aggregation Block\label{fig:img2}}{
        \includegraphics[width=0.9\linewidth]{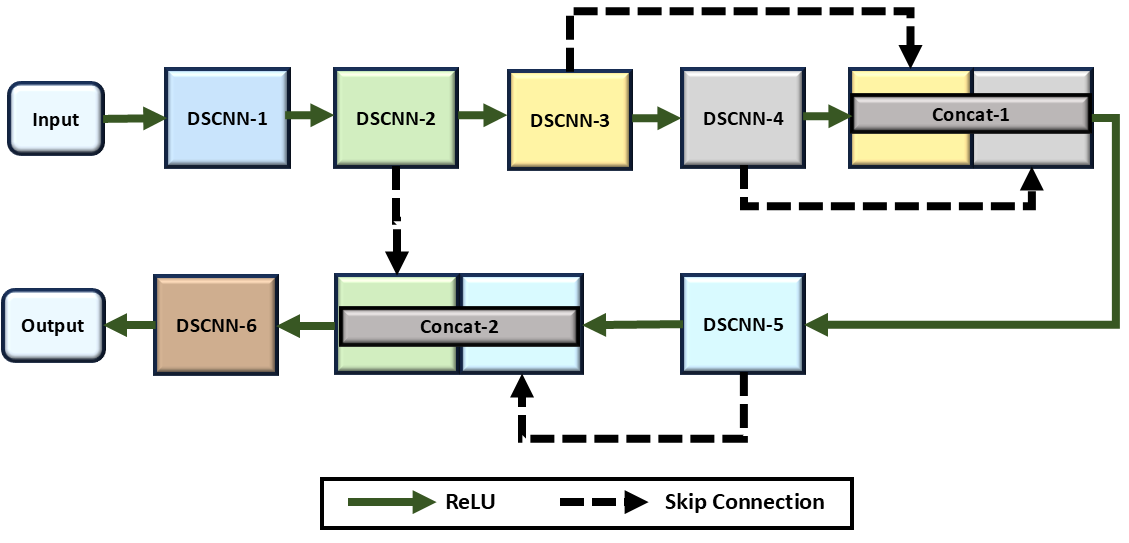}
    }\end{subcaptionbox}

    \vspace{0.3cm}

    \begin{subcaptionbox}{Spatial Attention\label{fig:img3}}{
        \includegraphics[width=0.7\linewidth]{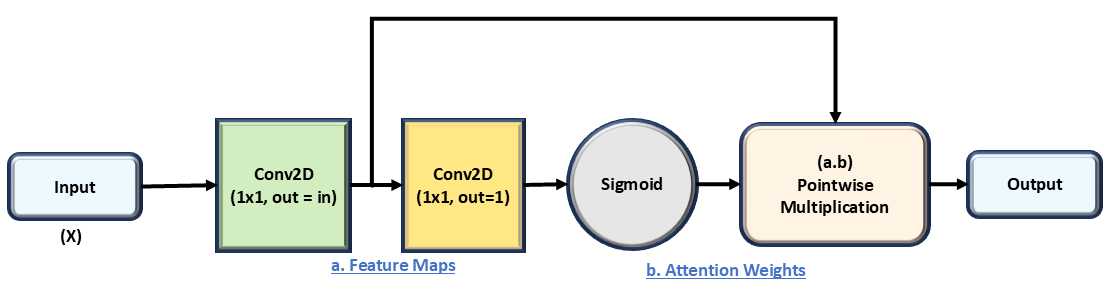}
    }\end{subcaptionbox}

    \caption{The Architecture of proposed \emph{LucentVisionNet} curve estimation part}
    \label{fig:three_imagesarchi}
\end{figure}

In this study, we introduce a novel image enhancement model that utilizes Depthwise Separable Convolutional Neural Networks (DSCNN) in conjunction with Spatial Attention. The initial step involves providing a comprehensive overview of the architectural elements of the model, encompassing the DSCNN blocks and the Spatial Attention module. Subsequently, we present the mathematical expressions for each constituent in order to elaborate on the functioning of the model. The proposed model is inspired by existing ZSL algorithms \cite{[47], Zero-DCE++, ZeroDCEsemantic}. The architecture for the proposed model is reported in Figure \ref {fig:three_imagesarchi}.\\
The image enhancement model proposed in this work is referred to as \emph{LucentVisionNet}. The proposed framework comprises three fundamental components: the Feature Extraction and Aggregation Block, the Spatial Attention and Deep Spatial Curve Estimation Network, and the Residual Learning module. The proposed model has been specifically developed to enhance low-light image quality through a perceptually-aware enhancement strategy. Additionally, the enhancement process is further refined through the utilization of residual learning techniques.

\subsection{Feature Extraction and Aggregation Block}\label{FE}
This block consists of a depth-wise convolution and a point-wise convolution that leads to Depthwise Separable Convolution. 
\subsubsection{Depthwise Separable Convolution Block}\label{DSCNN}
Depthwise separable convolution is an efficient alternative to traditional convolution operations, reducing computational cost while preserving performance. It consists of two primary stages: depthwise convolution and pointwise convolution \cite{howard2017mobilenets,chollet2017xception}.\\
\paragraph{Depthwise Convolution}
In the depthwise convolution step, each input channel is convolved independently with a separate 2D filter. This operation captures spatial features for each channel individually, significantly reducing computational complexity.\\
Let $\mathbf{X} \in \mathbb{R}^{C_{\text{in}} \times H \times W}$ represent the input feature map, where $C_{\text{in}}$ is the number of input channels, and $H$ and $W$ are the spatial dimensions. For each input channel $\mathbf{X}_c \in \mathbb{R}^{H \times W}$, a depthwise convolution is applied using a filter $\mathbf{W}_c \in \mathbb{R}^{K \times K}$:
\begin{equation}
\text{DWConv}(\mathbf{X}_c) = \mathbf{X}_c * \mathbf{W}_c,
\end{equation}
where $*$ denotes the 2D convolution operation. This process results in $C_{\text{in}}$ separate output feature maps.
\paragraph{Pointwise Convolution}
Following the depthwise step, pointwise convolution is applied using a $1 \times 1$ convolutional filter to combine the individual feature maps across channels~\cite{howard2017mobilenets}. Let $\mathbf{Y} \in \mathbb{R}^{C_{\text{in}} \times H \times W}$ denote the output of the depthwise convolution. A pointwise convolution filter $\mathbf{P} \in \mathbb{R}^{C_{\text{out}} \times C_{\text{in}} \times 1 \times 1}$ is applied as:
\begin{equation}
\text{PWConv}(\mathbf{Y}) = \mathbf{Y} * \mathbf{P},
\end{equation}
yielding an output feature map $\mathbf{Z} \in \mathbb{R}^{C_{\text{out}} \times H \times W}$, where each spatial location is a linear combination of all input channels.
\paragraph{Overall Depthwise Separable Convolution}
The combination of depthwise and pointwise convolutions defines the depthwise separable convolution~\cite{chollet2017xception}. It efficiently factorizes the standard convolution into a spatial convolution (depthwise) and a channel mixing operation (pointwise):
\begin{align}
\mathbf{Y} &= [\text{DWConv}(\mathbf{X}_1), \text{DWConv}(\mathbf{X}_2), \dots, \text{DWConv}(\mathbf{X}_{C_{\text{in}}})] \in \mathbb{R}^{C_{\text{in}} \times H \times W}, \\
\mathbf{Z} &= \text{PWConv}(\mathbf{Y}) \in \mathbb{R}^{C_{\text{out}} \times H \times W}.
\end{align}
This approach offers a substantial reduction in the number of parameters and computations, making it highly suitable for deployment in resource-constrained environments.
\subsection{Spatial Attention Block}\label{SA}
The spatial attention mechanism enhances the representational power of convolutional neural networks by assigning importance to different spatial regions in the input feature maps~\cite{woo2018cbam}. Given an input tensor $\mathbf{X} \in \mathbb{R}^{C_{\text{in}} \times H \times W}$, where $C_{\text{in}}$ is the number of input channels and $H \times W$ denotes the spatial resolution, the spatial attention block\cite{MIRNET} proceeds as follows:
\paragraph{Feature Map Projection.} First, the input tensor is projected using a $1 \times 1$ convolutional layer to generate intermediate feature maps:
\begin{equation}
\mathbf{F} = \mathbf{X} * \mathbf{W}_{\text{conv}},
\end{equation}
where $*$ denotes the 2D convolution operation, and $\mathbf{W}_{\text{conv}}$ represents the $1 \times 1$ convolutional kernel. The resulting tensor $\mathbf{F} \in \mathbb{R}^{C' \times H \times W}$ contains refined features from the input.
\paragraph{Attention Map Generation.} A second $1 \times 1$ convolutional layer is applied to $\mathbf{F}$ to generate a spatial attention map:
\begin{equation}
\mathbf{M} = \mathbf{F} * \mathbf{W}_{\text{att}},
\end{equation}
where $\mathbf{W}_{\text{att}}$ is another $1 \times 1$ convolutional kernel. The output $\mathbf{M} \in \mathbb{R}^{1 \times H \times W}$ represents the unnormalized attention weights over spatial dimensions.
\paragraph{Normalization via Sigmoid.} To ensure interpretability and constrain the attention values between 0 and 1, a sigmoid activation function $\sigma(\cdot)$ is applied:
\begin{equation}
\mathbf{A} = \sigma(\mathbf{M}).
\end{equation}
\paragraph{Attention-Weighted Output.} The final output is obtained by performing element-wise multiplication between the attention map $\mathbf{A}$ and the intermediate feature maps $\mathbf{F}$:
\begin{equation}
\mathbf{O} = \mathbf{F} \odot \mathbf{A},
\end{equation}
where $\odot$ denotes element-wise multiplication. This operation emphasizes informative spatial regions while suppressing less relevant ones.
This spatial attention mechanism improves the model’s ability to focus on significant spatial features, making it beneficial for tasks such as image classification and segmentation~\cite{woo2018cbam}.
\subsection{Multi-Scale Spatial Curve Estimation Network}
To capture both fine-grained details and high-level contextual information, the proposed architecture employs a multi-resolution feature extraction strategy. The input image $X \in \mathbb{R}^{C_{\text{in}} \times H \times W}$ is processed at three distinct resolutions: the original scale $X$, a half-scale downsampled version $X/2 \in \mathbb{R}^{C_{\text{in}} \times H/2 \times W/2}$, and a quarter-scale version $X/4 \in \mathbb{R}^{C_{\text{in}} \times H/4 \times W/4}$. These multi-resolution representations are independently fed into parallel \textbf{Feature Extraction and Aggregation Blocks \ref{FE}}, each denoted by different colors in Figure \ref{fig:three_imagesarchi} (green, blue, and yellow).\\
Each block is composed of a stack of \textbf{Depthwise Separable Convolutional Neural Networks (DSCNNs) \ref{DSCNN}}, where the $i$-th layer is denoted as $DWConv_i$. The operation of the DSCNN is defined as:
\begin{equation}
DWConv_i(X) = \text{ReLU}\left((X * K_{\text{depthwise}}^i) * K_{\text{pointwise}}^i\right),
\end{equation}
where $K_{\text{depthwise}}^i$ represents a $3 \times 3$ depthwise convolution kernel applied separately to each input channel, and $K_{\text{pointwise}}^i$ is a $1 \times 1$ convolution kernel used to combine the resulting outputs. This design significantly reduces computational complexity while preserving critical spatial and semantic features.\\
The outputs of the feature extraction modules at each resolution are denoted as $D_1$, $D_2$, and $D_3$ corresponding to input scales $X$, $X/2$, and $X/4$, respectively.\\
Following the extraction stage, multi-scale outputs undergo a comprehensive fusion process that includes \textit{upsampling}, \textit{feature aggregation}, \textit{spatial attention}, and \textit{final prediction}.\\
To ensure uniform spatial dimensions, the outputs $D_2$ and $D_3$ are upsampled by factors of 2 and 4, respectively, aligning them with the resolution of $D_1$. The fusion is conducted through concatenation followed by additional DSCNN layers, facilitating hierarchical integration of features. The aggregation process is formally represented as:
\begin{align}
F_1 &= \text{DSCNN}\left(\text{Concat}(D_1, \text{Upsample}(D_2))\right), \\
F_2 &= \text{DSCNN}\left(\text{Concat}(F_1, \text{Upsample}(D_3))\right), \\
F_{\text{agg}} &= \text{Concat}(F_1, F_2).
\end{align}
This hierarchical fusion strategy ensures the effective integration of local and global contextual information. The final aggregated feature map $F_{\text{agg}}$ is then processed through a \textbf{Spatial Attention Block\ref{SA}}, which adaptively enhances important spatial regions while suppressing less relevant areas.\\
The attention-refined features are subsequently passed through a concluding \textbf{DSCNN layer} followed by a \textbf{Tanh activation function} to generate the final prediction map. This output stage, labeled as \texttt{Conv\_Final} in Figure~\ref{fig:three_imagesarchi}, transforms the fused features into the desired output space.
This end-to-end architectural design efficiently explores multiscale contextual information and spatial saliency while maintaining low computational overhead, making it particularly effective for applications requiring precise spatial understanding.
\subsection{Residual Learning}
\begin{figure}
    \centering
    \includegraphics[width=\linewidth]{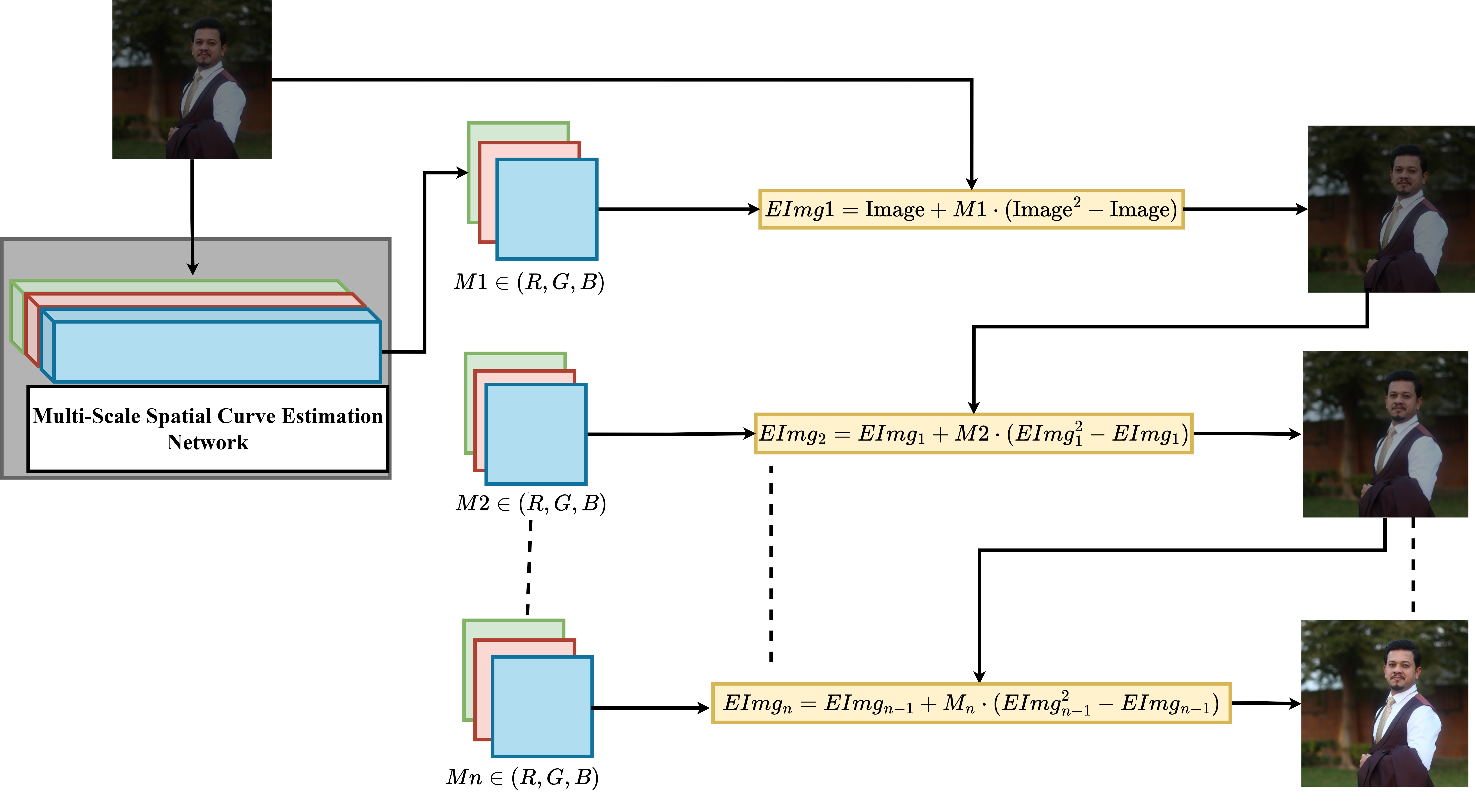}
    \caption{Residual Learning based architecture for \emph{LucentVisionNet}}
    \label{fig:lucent}
\end{figure}

Our proposed method leverages the benefits of residual learning inspired by contemporary ZSL architectures~\cite{[47], Zero-DCE++, ZeroDCEsemantic} to improve the image enhancement procedure. The incorporation of residual connections offers distinct advantages that substantially impact both the efficacy and efficiency of our strategy. These advantages span various dimensions, including gradient propagation, vanishing gradient mitigation, preservation of image particulars, and the handling of complex mappings.\\
The inclusion of residual connections ensures efficient gradient flow during training, expediting convergence, and preventing gradient degradation. The pipeline of the proposed method is illustrated in Figure~\ref{fig:lucent}.
The mathematical formulation for the recurrent quadratic equation-based learning process\cite{[47]} is expressed as follows:
\begin{equation}
X_t = X_{t-1} + D \left( X_{t-1}^2 - X_{t-1} \right)
\end{equation}
\noindent where:
\begin{itemize}
    \item $X_t$ represents the enhanced image at iteration $t$,
    \item $X_{t-1}$ is the image from the previous iteration,
    \item $D$ is a diagonal matrix with enhancement factors $x_r$ as its diagonal elements.
\end{itemize}
The diagonal matrix $D$ used in this matrix-based formulation effectively mitigates the problems of vanishing gradients, allowing the network to learn deeper representations with greater stability. The distinctive inclusion of a residual term in each iteration helps preserve critical image details, allowing the network to refine subtle features while maintaining the structural integrity of the image. \\
This design choice, derived from the principles of residual learning, enhances the network's ability to capture complex nonlinear mappings, thereby increasing its adaptability to diverse enhancement requirements. Collectively, these properties contribute to a robust image enhancement framework that improves visual quality while preserving the authenticity of the original content.
\section{Loss Function}
The composite loss function used to train the enhancement network is a weighted combination of multiple complementary losses, each designed to guide the network toward generating perceptually high-quality, naturally illuminated, and semantically consistent images. The composite loss is formulated as:
\begin{equation}
\mathcal{L}_{\text{composite}} = \lambda_{\text{TV}} \mathcal{L}_{\text{TV}}(A) + \mathcal{L}_{\text{spa}}(I_{\text{enh}}, I_{\text{low}}) + \lambda_{\text{col}} \mathcal{L}_{\text{color}}(I_{\text{enh}}) + \lambda_{\text{exp}} \mathcal{L}_{\text{exp}}(I_{\text{enh}}, E) + \lambda_{\text{seg}} \mathcal{L}_{\text{seg}}(I_{\text{enh}}) + \lambda_{\text{NR}} \mathcal{L}_{\text{NR}}(I_{\text{enh}})
\end{equation}
Where:
\begin{itemize}
    \item \( I_{\text{enh}} \) is the enhanced image,
    \item \( I_{\text{low}} \) is the low-light input image,
    \item \( A \) is the learned enhancement map,
    \item \( E \) is the reference exposure map,
    \item \( \lambda_{\text{TV}} = 1600 \), \( \lambda_{\text{col}} = 5 \), \( \lambda_{\text{exp}} = 10 \), \( \lambda_{\text{seg}} = 0.1 \), and \( \lambda_{\text{NR}} = 0.1 \) are weighting factors.
\end{itemize}
Each component of the composite loss is described below:
\subsection{Total Variation Loss (\( \mathcal{L}_{\text{TV}} \))}
This loss encourages spatial smoothness in the enhancement map \( A \), preventing abrupt changes and noise:
\begin{equation}
\mathcal{L}_{\text{TV}}(A) = \sum_{i,j} \left( \left\| A_{i,j+1} - A_{i,j} \right\|^2 + \left\| A_{i+1,j} - A_{i,j} \right\|^2 \right)
\end{equation}
\subsection{Spatial Consistency Loss \texorpdfstring{$\mathcal{L}_{\text{spa}}$}{Lspa}}
To preserve local structural consistency between the original image and the enhanced output, we define a \emph{Spatial Consistency Loss} that constrains the directional gradients of the enhanced image to remain close to those of the input image.
Let \( I \) be the original RGB input image and \( \hat{I} \) be the enhanced image. Both are first converted to grayscale by channel-wise averaging:
\begin{equation}
I_{\text{gray}} = \frac{1}{3} \sum_{c=1}^{3} I_c, \quad 
\hat{I}_{\text{gray}} = \frac{1}{3} \sum_{c=1}^{3} \hat{I}_c
\end{equation}
We then apply an average pooling operation \( \mathcal{P}(\cdot) \) with kernel size \( 4 \times 4 \) to reduce noise:
\begin{equation}
I_p = \mathcal{P}(I_{\text{gray}}), \quad 
\hat{I}_p = \mathcal{P}(\hat{I}_{\text{gray}})
\end{equation}
Directional gradients are computed using four fixed convolutional kernels \( K_d \) corresponding to directions \( d \in \{\text{left}, \text{right}, \text{up}, \text{down}\} \). The directional gradients are given by:
\begin{equation}
\nabla_d I = I_p * K_d, \quad 
\nabla_d \hat{I} = \hat{I}_p * K_d
\end{equation}
where \( * \) denotes convolution. The spatial consistency loss is defined as the sum of squared differences of directional gradients:
\begin{equation}
\mathcal{L}_{\text{spa}} = 
\sum_{d \in \{\text{left}, \text{right}, \text{up}, \text{down}\}} 
\left\| \nabla_d I - \nabla_d \hat{I} \right\|_2^2
\end{equation}
Explicitly, this can be written as:
\begin{align}
\mathcal{L}_{\text{spa}} = 
& \left\| \nabla_{\text{left}} I - \nabla_{\text{left}} \hat{I} \right\|_2^2 
+ \left\| \nabla_{\text{right}} I - \nabla_{\text{right}} \hat{I} \right\|_2^2 \nonumber \\
+ & \left\| \nabla_{\text{up}} I - \nabla_{\text{up}} \hat{I} \right\|_2^2 
+ \left\| \nabla_{\text{down}} I - \nabla_{\text{down}} \hat{I} \right\|_2^2
\end{align}
This loss enforces structural similarity between the input and enhanced images, particularly in local edge and texture regions.
\subsection{Color Constancy Loss (\( \mathcal{L}_{\text{color}} \))}
Encourages realistic color balance by minimizing deviation between the RGB channels:
\begin{equation}
\mathcal{L}_{\text{color}} = \sqrt{(R - G)^2 + (R - B)^2 + (G - B)^2}
\end{equation}
where \( R, G, B \) denote the mean intensities of the red, green, and blue channels of the enhanced image.
\subsection{Exposure Control Loss (\( \mathcal{L}_{\text{exp}} \))}
Regulates the exposure level of the enhanced image toward a reference exposure map \( E \):
\begin{equation}
\mathcal{L}_{\text{exp}} = \left\| \text{AvgPool}(I_{\text{enh}}) - E \right\|^2
\end{equation}
\subsection{Segmentation Guidance Loss (\( \mathcal{L}_{\text{seg}} \))}
This auxiliary loss promotes semantic fidelity by penalizing deviations in segmentation structure between the enhanced image and a reference segmentation map, typically using an unsupervised segmentation network \cite{ZeroDCEsemantic}.
\subsection{No-Reference Image Quality Loss \texorpdfstring{$\mathcal{L}_{\text{NR}}$}{LNR}}
Using the MUSIQ-AVA model, we apply a No-Reference Image Quality Assessment (NR-IQA) loss to guarantee that the improved image is perceptually high-quality from a human perspective \cite{musiq}.  This model was trained using the AVA dataset \cite{ava}, which includes aesthetic quality annotations from human assessments, and is based on the Multiscale Image Quality Transformer (MUSIQ) architecture.\\
Let \( \hat{I} \) denote the enhanced image. The MUSIQ-AVA model predicts an aesthetic quality score \( S(\hat{I}) \in [0, 100] \), where higher values correspond to higher perceptual quality. We define the no-reference loss as the deviation from the maximum possible aesthetic score:
\begin{equation}
\mathcal{L}_{\text{NR}} = 100 - \mathbb{E}\left[ S(\hat{I}) \right]
\end{equation}
where \( \mathbb{E}[ \cdot ] \) denotes the mean over the batch of predicted quality scores. This formulation encourages the enhancement network to generate images that maximize the perceived quality.\\
The MUSIQ-AVA model supports gradient backpropagation, allowing it to be used directly as a loss function:
\begin{itemize}
    \item The model is instantiated with \texttt{as\_loss=True} to enable its use in training.
    \item During the forward pass, the aesthetic score is computed and its mean is taken across the batch.
    \item The loss is then defined as the difference between the maximum quality score (100) and the average score.
\end{itemize}
This no-reference loss is especially important in scenarios where ground-truth high-quality images are unavailable, and subjective perceptual quality becomes a key optimization criterion, and so used in the current work.
\subsection{Final Objective}
\begin{equation}
\mathcal{L}_{\text{composite}} = 1600 \cdot \mathcal{L}_{\text{TV}} + \mathcal{L}_{\text{spa}} + 5 \cdot \mathcal{L}_{\text{color}} + 10 \cdot \mathcal{L}_{\text{exp}} + 0.1 \cdot \mathcal{L}_{\text{seg}} + 0.1 \cdot \mathcal{L}_{\text{NR}}
\end{equation}
This composite loss ensures that the enhanced outputs are perceptually natural, well-exposed, structurally faithful, and aesthetically pleasing.
\section{Experimental Settings}
\subsection{Implementation Details}
In alignment with Zero-DCE \cite{[47], Zero-DCE++, ZeroDCEsemantic}, our training strategy leverages a dataset specifically curated to include both low-light and over-exposed conditions, enabling the model to learn dynamic range enhancement effectively. In particular, we use 360 multi-exposure sequences from the Part1 subset of the SICE dataset \cite{SICE}.  We extract a total of 3,022 images with varying exposure settings from them. Consistent with prior work such as EnlightenGAN \cite{[48]}, we randomly split the dataset into 2,422 images for training and 600 for validation. All images are resized to \(512 \times 512 \times 3\) to maintain consistency during training and evaluation.\\
This training configuration ensures robustness for real-world low-light and overexposed image enhancement tasks by improving the model's generalization across a range of illumination conditions.  The trained model is evaluated on multiple subsets of different datasets selected from earlier research in the same field in order to test the proposed approach in real time.  Table \ref{tab:testsets_comparison} contains the specifics of these test sets.
\begin{table}[htbp]
\centering
\caption{Comparison of test sets used for evaluation of performance and comparative analysis}
\begin{tabular}{|l|c|c|c|p{5cm}|}
\hline
\textbf{Dataset} & \textbf{\#Images} & \textbf{Paired} & \textbf{Real/Synthetic} & \textbf{Description} \\
\hline
\textbf{DarkBDD} \cite{ZeroDCEsemantic} & 100 & No & Real & 100 lowlight images selected from the orignal BDD10K\cite{yu2020bdd100k} and created a new test set called as DarkBDD\cite{ZeroDCEsemantic} \\
\hline
\textbf{DarkCityScape} \cite{ZeroDCEsemantic} & 150 & No & Real & Gamma correction applied on selected 150 images from original CityScape\cite{cordts2016cityscapes} dataset to create DarkCityScape\cite{ZeroDCEsemantic} \\
\hline
\textbf{DICM} \cite{lee2012contrast} & 69 & No & Real & Diverse indoor and outdoor low-light images captured with commercial digital cameras. \\
\hline
\textbf{LIME} \cite{guo2016lime} & 10 & No & Real & Real-world low-light images. \\
\hline
\textbf{LOL} \cite{wei2018deep} & 15 & Yes & Real/Synthetic & Paired widely used benchmark for supervised enhancement methods. Selected 15 images. \\
\hline
\textbf{LOL-v2 } \cite{yang2021fidelity} & 100+100 & Yes & Real & Improved version with more diverse scenes and better alignment; used 100 testing images for both real and synthetic. \\
\hline
\textbf{MEF} \cite{ma2015perceptual} & 17 & No & Real & Multi-exposure fusion sequences. \\
\hline
\textbf{NPE} \cite{wang2013naturalness} & 8 & No & Real & Nighttime photographs with varied exposure levels. \\
\hline
\textbf{VV} \cite{VV} & 24 & No & Real & Night road scenarios; used in autonomous driving and low-light enhancement studies. \\
\hline
\end{tabular}
\label{tab:testsets_comparison}
\end{table}
\subsection{Performance Validation Metrics}
Peak Signal-to-Noise Ratio (PSNR), Structural Similarity Index Measure (SSIM), Feature Similarity Index Measure (FSIM), Visual Saliency-Induced Index (VSI), Learned Perceptual Image Patch Similarity (LPIPS), Deep Image Structure and Texture Similarity (DISTS), and Mean Absolute Difference (MAD) \cite{wang2004image, hore2010image, zhang2011fsim, zhang2014vsi, zhang2018unreasonable, ding2020image}. were among the full-reference image quality assessment metrics used to thoroughly assess the performance of the proposed approach. Since these metrics are good at measuring distortion in comparison to a reference ground truth image, they are frequently utilized in the field of image quality assessment. Their reliance on the availability of reference images, however, is a major drawback, especially in situations involving image augmentation in the real world where a ground truth might be arbitrary or nonexistent.\\
To address this limitation, the study integrates no-reference or blind image quality assessment (BIQA) models that do not require a reference image. These include advanced learning-based approaches such as NIMA \cite{NIMA}, PaQ2PiQ \cite{PaQ2PiQ}, DBCNN \cite{DBCNN}, MUSIQ-Koniq \cite{musiq}, MANIQA \cite{Yang2022MANIQA}, CLIP-IQA \cite{Wang2023CLIPIQA}, TReS-Koniq \cite{Lin2022TReS}, HyperIQA \cite{Su2020HyperIQA}, GPR-BIQA \cite{khalid2021gaussian}, QualityNet \cite{aslam2024qualitynet}, and PIQI \cite{ahmed2021piqi}. These BIQA techniques are particularly valuable in enhancement tasks, which often yield images that are plausible variants rather than exact replicas of an assumed "ideal" reference. In such cases, blind assessment methods offer a more contextually appropriate and perceptually aligned evaluation of image quality \cite{mittal2012no, zhang2021blind, ahmed2020perceptual}.
\section{Experimental Results}
The experimental validation of the performance of the proposed method is organized into two primary subsections: (1) qualitative assessment through visual comparisons and (2) quantitative evaluation using performance metrics. For a comprehensive comparative analysis, the proposed algorithm is evaluated against both paired and unpaired image enhancement methods. The paired methods include BIMEF \cite{BIMEF}, LIME \cite{LIME}, MF \cite{MF}, and Multiscale Retinex \cite{MultiscaleRetinex}, which are based on corresponding ground truth images during training. In contrast, the unpaired methods such as EnlightenGAN \cite{jiang2021enlightengan}, Zero-DCE \cite{[47]}, Zero-DCE++ \cite{Zero-DCE++}, and the Semantic-Guided Zero-Shot Learning framework \cite{ZeroDCEsemantic} operate without ground truth references, making them more suitable for real-world scenarios. This two-fold comparison enables a robust evaluation of the generalizability and effectiveness of the proposed method across different learning paradigms.
\subsection{Qualitative assessment through visual comparison}\label{qu}
\begin{figure}[htbp]
    \centering
    \includegraphics[width=\linewidth]{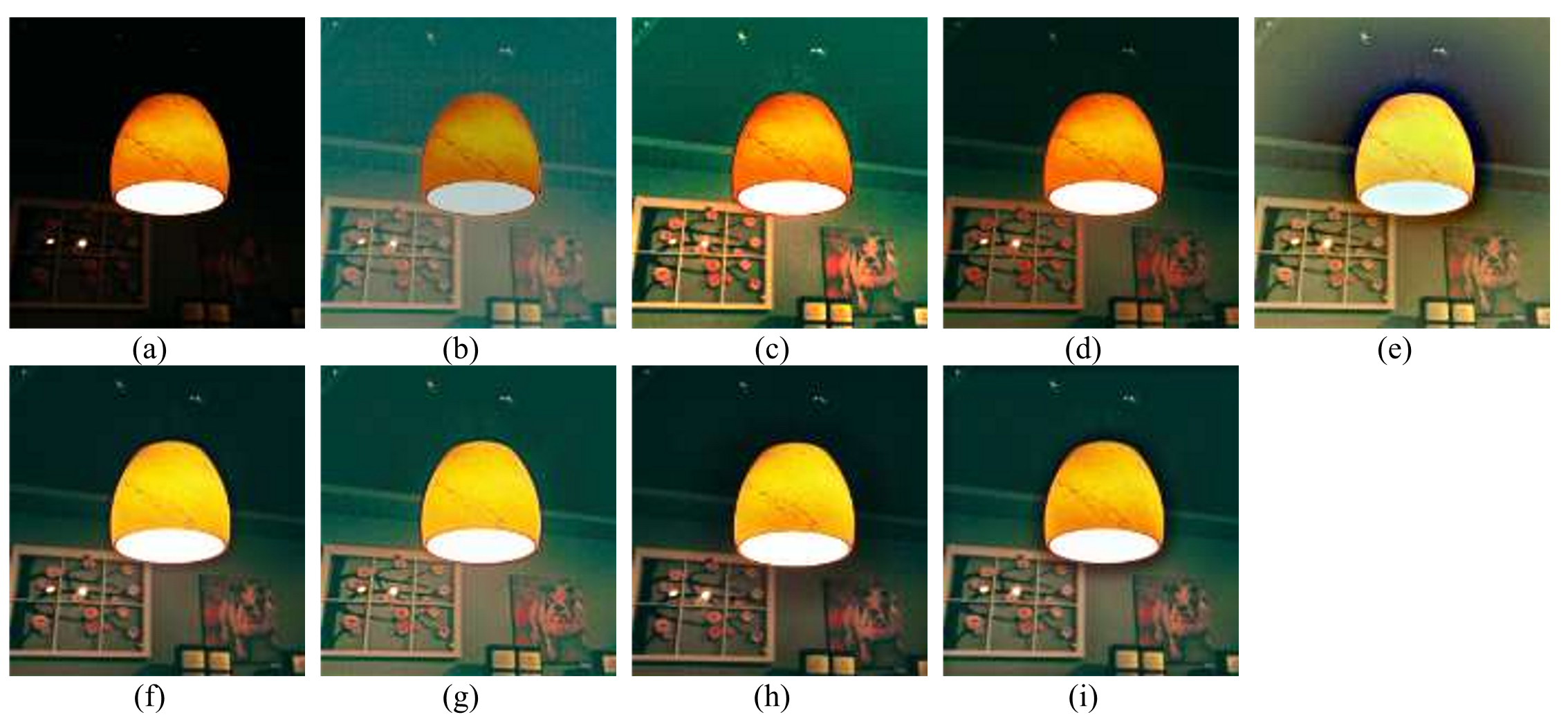}
    \caption{(a) Dark, (b) BIMEF\cite{BIMEF}, (c) LIME\cite{LIME}, (d) MF\cite{MF}, (e) Multiscale Retenix\cite{MultiscaleRetinex}, (f) ZERODCE\cite{[47]}, (g) ZERODCE$++$\cite{Zero-DCE++}, (h) Semantic Guide ZERO DCE\cite{ZeroDCEsemantic} and (h) Ours on LIME dataset}
    \label{fig:compLIME}
\end{figure}

\begin{figure}[!htb]
    \centering
    \includegraphics[width=\linewidth]{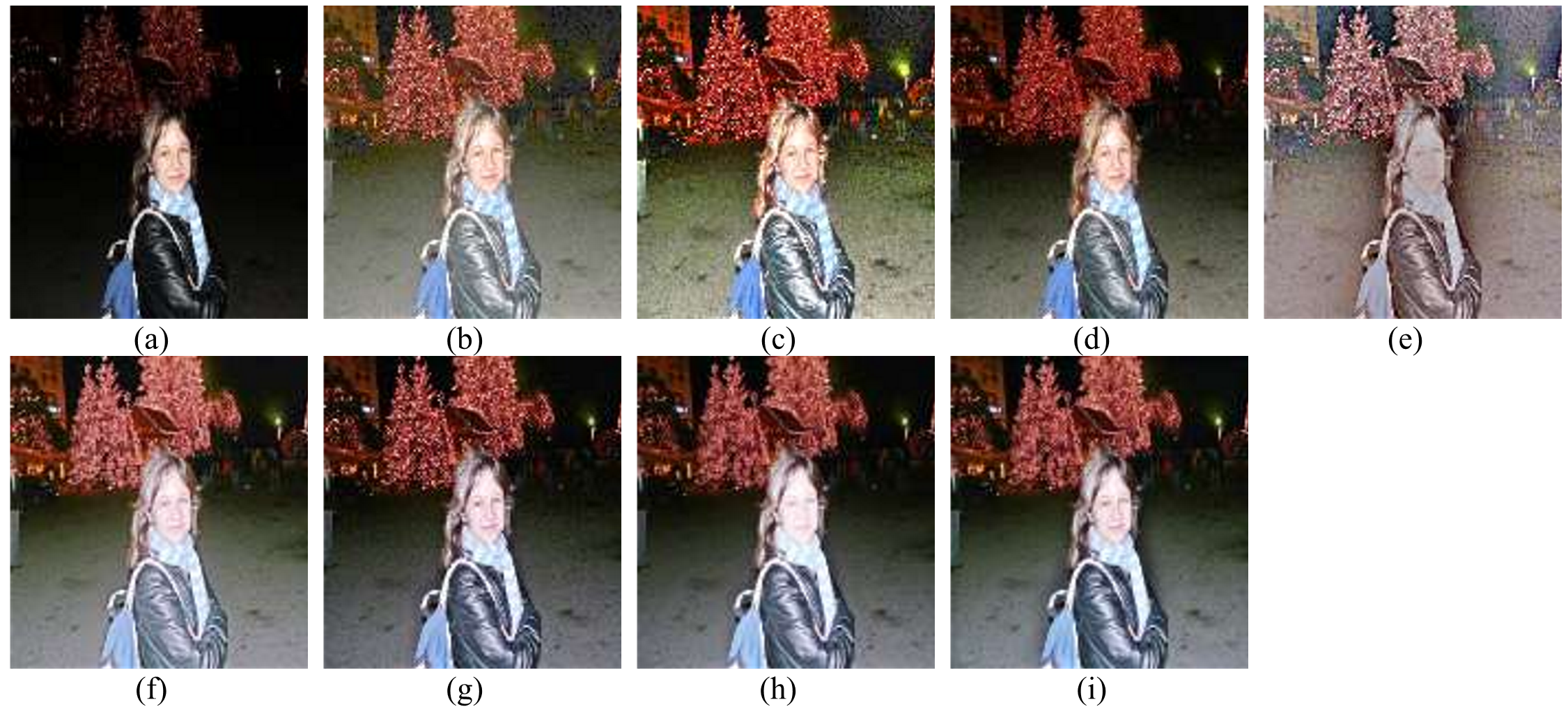}
    \caption{(a) Dark, (b) BIMEF\cite{BIMEF}, (c) LIME\cite{LIME}, (d) MF\cite{MF}, (e) Multiscale Retenix\cite{MultiscaleRetinex}, (f) ZERODCE\cite{[47]}, (g) ZERODCE$++$\cite{Zero-DCE++}, (h) Semantic Guide ZERO DCE\cite{ZeroDCEsemantic} and (h) Ours on VV dataset}
    \label{fig:compVV}
\end{figure}

Figure~\ref{fig:compLIME} presents the comparative analysis of a representative sample from the LIME dataset, while Figure~\ref{fig:compVV} illustrates the corresponding results for the VV dataset. The visual outcomes demonstrate that the proposed model performs adaptive image enhancement, successfully preserving the original color balance, contrast levels, and perceptual fidelity. This suggests the model's effectiveness in producing visually pleasing and structurally consistent enhancements across varying image conditions.
\subsubsection{Quantitative Evaluation using Performance Metrics for Unpaired Datasets}
Table~\ref{tab:iqabdd} presents a comprehensive quantitative comparison of various low-light image enhancement techniques on the DarkBDD dataset using multiple no-reference image quality metrics. While traditional paired-supervision methods like Multiscale Retinex and MF achieve competitive scores in metrics such as NIMA, DBCNN, and QualityNet, and EnlightenGAN (unsupervised) performs well in TReS-Koniq, zero-shot approaches show a promising balance between performance and generalizability. Among them, our proposed method consistently outperforms all others, achieving the highest average score (18.06) and leading across several key metrics including TReS-Koniq (49.08), QualityNet (0.72), and PIQI (0.70). These results highlight the effectiveness of our model in enhancing low-light driving scenes without requiring paired data or task-specific training. In addition on average our algorithm outperforms the other low-light enhancement algorithms. 
\begin{table}[htbp]
\centering
\caption{Comparison of no-reference image quality metrics for various enhancement techniques on the DarkBDD~\cite{ZeroDCEsemantic} dataset}
\label{tab:iqabdd}
\scriptsize
\begin{adjustbox}{width=\textwidth}
\begin{tabular}{lcccc|c|cccc}
\toprule
\multirow{3}{*}{\textbf{Metric}} &
\multicolumn{4}{c|}{\textbf{Paired Supervision}} &
\multicolumn{1}{c|}{\textbf{Unsupervised}} &
\multicolumn{4}{c}{\textbf{Zero-Shot Learning}} \\
\cmidrule(lr){2-5} \cmidrule(lr){6-6} \cmidrule(lr){7-10}
& BIMEF\cite{BIMEF} & LIME\cite{LIME} & MF\cite{MF} & Multiscale Retinex\cite{MultiscaleRetinex} & EnlightenGAN\cite{jiang2021enlightengan} & SemZSL\cite{ZeroDCEsemantic} & Zero-DCE\cite{[47]} & Zero-DCE++\cite{Zero-DCE++} & Ours \\
\midrule
NIMA\cite{NIMA}         & 3.89  & 3.88  & 3.85  & 4.13  & 3.87  & 3.86  & 3.62  & 3.64  & 3.99 \\
PaQ2PiQ\cite{PaQ2PiQ}      & 66.99 & 68.55 & 68.30 & 67.24 & 66.38 & 66.37 & 67.31 & 67.33 & 66.73 \\
DBCNN\cite{DBCNN}        & 31.98 & 31.98 & 32.00 & 33.76 & 32.87 & 33.02 & 29.85 & 29.87 & 33.46 \\
MUSIQ-Koniq\cite{musiq}  & 45.92 & 41.92 & 43.77 & 41.29 & 41.89 & 41.76 & 39.16 & 39.18 & 42.30 \\
MANIQA\cite{Yang2022MANIQA}       & 0.59  & 0.56  & 0.58  & 0.54  & 0.55  & 0.55  & 0.53  & 0.55  & 0.55 \\
CLIPIQA\cite{Wang2023CLIPIQA}      & 0.14  & 0.12  & 0.13  & 0.13  & 0.13  & 0.13  & 0.12  & 0.14  & 0.13 \\
TReS-koniq\cite{Lin2022TReS}   & 45.61 & 46.04 & 47.46 & 47.77 & 47.84 & 47.79 & 41.50 & 41.52 & 49.08 \\
HyperIQA\cite{Su2020HyperIQA}     & 0.30  & 0.30  & 0.29  & 0.28  & 0.30  & 0.30  & 0.27  & 0.29  & 0.30 \\
GPR-BIQA\cite{khalid2021gaussian}     & 0.58  & 0.60  & 0.59  & 0.57  & 0.58  & 0.59  & 0.56  & 0.58  & 0.65 \\
Quality Net\cite{aslam2024qualitynet}  & 0.60  & 0.64  & 0.66  & 0.63  & 0.65  & 0.64  & 0.61  & 0.63  & 0.72 \\
PIQI\cite{ahmed2021piqi}         & 0.62  & 0.63  & 0.64  & 0.60  & 0.61  & 0.63  & 0.59  & 0.61  & 0.70 \\
\midrule
\textbf{Average} & 17.93 & 17.75 & 18.02 & 17.90 & 17.79 & 17.79 & 16.74 & 16.76 & \textbf{18.06} \\
\bottomrule
\end{tabular}
\end{adjustbox}
\end{table}
The results summarized in Table~\ref{tab:cityscape_quality} demonstrate a clear performance margin of our proposed method over existing enhancement techniques across a comprehensive set of no-reference image quality metrics. While traditional paired-supervised algorithms (e.g., BIMEF, LIME, Retinex) show competitive results in selected metrics, their generalization to unstructured, real-world inputs remains limited. Unsupervised models like EnlightenGAN and zero-shot methods such as Zero-DCE and Zero-DCE++ offer a more flexible training paradigm, yet their performance remains suboptimal in several perceptual and deep feature-based assessments (e.g., DBCNN, TReS-Koniq, HyperIQA). In contrast, our approach yields superior average performance (24.61), indicating robust enhancement capability and perceptual fidelity. This highlights the effectiveness of our model in learning meaningful representations without the need for explicit paired supervision, making it highly suitable for real-world applications in automated visual systems, especially where ground-truth data is scarce or unavailable.
\begin{table*}[htbp]
\centering
\caption{Comparison of No-Reference Image Quality Metrics for Various Enhancement Techniques on the DarkCityScape\cite{ZeroDCEsemantic} Dataset}
\label{tab:cityscape_quality}
\scriptsize
\begin{adjustbox}{width=\textwidth}
\begin{tabular}{lcccc|c|cccc}
\toprule
\multirow{3}{*}{\textbf{Metric}} &
\multicolumn{4}{c|}{\textbf{Paired Supervision}} &
\multicolumn{1}{c|}{\textbf{Unsupervised}} &
\multicolumn{4}{c}{\textbf{Zero-Shot Learning}} \\
\cmidrule(lr){2-5} \cmidrule(lr){6-6} \cmidrule(lr){7-10}
& BIMEF\cite{BIMEF} & LIME\cite{LIME} & MF\cite{MF} & Multiscale Retinex\cite{MultiscaleRetinex} & EnlightenGAN\cite{jiang2021enlightengan} & SemZSL\cite{ZeroDCEsemantic} & Zero-DCE\cite{[47]} & Zero-DCE++\cite{Zero-DCE++} & Ours \\
\midrule
NIMA\cite{NIMA}        & 4.43 & 4.70 & 4.60 & 4.77 & 4.70 & 4.70 & 4.41 & 4.43 &  4.75 \\
PaQ2PiQ\cite{PaQ2PiQ}     & 61.31 & 72.99 & 72.24 & 73.67 & 66.11 & 66.11 & 68.95 & 68.97 &  68.61 \\
DBCNN\cite{DBCNN}      & 43.89 & 55.13 & 54.61 & 54.82 & 57.12 & 57.12 & 45.84 & 45.86 &  58.68 \\
musiq-koniq\cite{musiq} & 50.79 & 61.31 & 61.33 & 60.43 & 57.58 & 57.58 & 56.97 & 56.99 &  60.41 \\
MANIQA\cite{Yang2022MANIQA}      & 0.60  & 0.59  & 0.59  & 0.58  & 0.61  & 0.61  & 0.57  & 0.59  &  0.61 \\
CLIPIQA\cite{Wang2023CLIPIQA}     & 0.20  & 0.18  & 0.17  & 0.20  & 0.25  & 0.25  & 0.16  & 0.18  &  0.23 \\
TReS-koniq\cite{Lin2022TReS}  & 56.83 & 64.90 & 67.58 & 66.58 & 72.28 & 72.25 & 58.72 & 58.74 &  74.69 \\
HyperIQA\cite{Su2020HyperIQA}     & 0.43  & 0.47  & 0.48  & 0.49  & 0.53  & 0.53  & 0.43  & 0.45  &  0.54 \\
GPR-BIQA\cite{khalid2021gaussian}    & 0.62  & 0.67  & 0.65  & 0.66  & 0.68  & 0.67  & 0.60  & 0.62  &  0.73 \\
Quality Net\cite{aslam2024qualitynet} & 0.65  & 0.69  & 0.67  & 0.68  & 0.71  & 0.70  & 0.63  & 0.65  &  0.75 \\
PIQI\cite{ahmed2021piqi}         & 0.64  & 0.68  & 0.67  & 0.69  & 0.70  & 0.69  & 0.62  & 0.64  &  0.74 \\
\midrule
\textbf{Average} & 20.04 & 23.85 & 23.96 & 23.96 & 23.75 & 23.75 & 21.63 & 21.65 &  \textbf{24.61} \\
\bottomrule
\end{tabular}
\end{adjustbox}
\end{table*}
The evaluation on the DICM dataset, summarized in Table~\ref{tab:dicm_quality}, reveals the superior performance of our proposed approach in comparison with both paired and unpaired supervision-based enhancement algorithms. Among paired methods such as BIMEF, LIME, and Multiscale Retinex, moderate performance was observed across most metrics, indicating their effectiveness under constrained settings but limited adaptability. Unsupervised and zero-shot learning techniques, including EnlightenGAN, Semantic Guided-ZSL, and Zero-DCE, yielded mixed results, often falling behind in perceptual and feature-based metrics such as MUSIQ-Koniq, TReS-Koniq, and DBCNN. Notably, our method achieved the highest average score (24.76), demonstrating its robustness and ability to maintain perceptual quality and structural integrity in diverse illumination conditions. This reinforces the generalizability and effectiveness of our model under real-world imaging scenarios where paired training data is unavailable.
\begin{table*}[htbp]
\centering
\caption{Comparison of No-Reference Image Quality Metrics for Various Enhancement Techniques on the DICM Dataset \cite{lee2012contrast}}
\label{tab:dicm_quality}
\scriptsize
\begin{adjustbox}{width=\textwidth}
\begin{tabular}{lcccc|c|cccc}
\toprule
\multirow{3}{*}{\textbf{Metric}} &
\multicolumn{4}{c|}{\textbf{Paired Supervision}} &
\multicolumn{1}{c|}{\textbf{Unsupervised}} &
\multicolumn{4}{c}{\textbf{Zero-Shot Learning}} \\
\cmidrule(lr){2-5} \cmidrule(lr){6-6} \cmidrule(lr){7-10} 
& BIMEF\cite{BIMEF} & LIME\cite{LIME} & MF\cite{MF} & Multiscale Retinex\cite{MultiscaleRetinex} & EnlightenGAN\cite{jiang2021enlightengan} & SemZSL\cite{ZeroDCEsemantic} & Zero-DCE\cite{[47]} & Zero-DCE++\cite{Zero-DCE++} & Ours \\
\midrule
NIMA\cite{NIMA}        & 4.52  & 4.42  & 4.56  & 4.58  & 4.29  & 4.29  & 4.22  & 4.24  & 4.36 \\
PaQ2PiQ\cite{PaQ2PiQ}      & 74.58 & 76.17 & 75.60 & 74.42 & 73.37 & 73.31 & 73.91 & 73.93 & 76.95 \\
DBCNN\cite{DBCNN}        & 51.36 & 50.58 & 51.60 & 52.89 & 45.61 & 45.95 & 44.53 & 44.55 & 52.38 \\
musiq-koniq\cite{musiq}  & 62.28 & 61.37 & 62.82 & 60.04 & 56.91 & 57.19 & 56.41 & 56.43 & 68.46 \\
MANIQA\cite{Yang2022MANIQA}       & 0.66  & 0.64  & 0.66  & 0.61  & 0.62  & 0.62  & 0.61  & 0.63  & 0.63 \\
CLIPIQA\cite{Wang2023CLIPIQA}      & 0.56  & 0.52  & 0.55  & 0.49  & 0.56  & 0.55  & 0.54  & 0.56  & 0.56 \\
TReS-koniq\cite{Lin2022TReS}  & 69.64 & 64.48 & 70.45 & 70.33 & 65.16 & 65.19 & 64.22 & 64.24 & 66.42\\
HyperIQA\cite{Su2020HyperIQA}     & 0.53  & 0.52  & 0.54  & 0.53  & 0.44  & 0.44  & 0.42  & 0.44  & 0.45\\
GPR-BIQA\cite{khalid2021gaussian}     & 0.70  & 0.68  & 0.69  & 0.67  & 0.61  & 0.61  & 0.60  & 0.62  & 0.72 \\
Quality Net\cite{aslam2024qualitynet}  & 0.72  & 0.70  & 0.65  & 0.69  & 0.63  & 0.63  & 0.62  & 0.63  & 0.74 \\
PIQI\cite{ahmed2021piqi}         & 0.71  & 0.69  & 0.64  & 0.68  & 0.62  & 0.62  & 0.61  & 0.60  & 0.73 \\
\midrule
Average      & 24.21 & 23.71 & 24.43 & 24.18 & 22.62 & 22.67 & 22.43 & 22.44 & \textbf{24.76} \\
\bottomrule
\end{tabular}
\end{adjustbox}
\end{table*}
Table~\ref{tab:lime_quality} presents a comparative analysis of various enhancement techniques based on no-reference image quality metrics evaluated on the LIME dataset. The metrics include traditional models such as \texttt{nima}, \texttt{paq2piq}, \texttt{dbcnn}, and advanced neural-based models like \texttt{musiq-koniq}, \texttt{maniqa}, \texttt{tres-koniq}, and \texttt{GPR-BIQA}. Among all the techniques, our proposed method consistently achieves superior performance across nearly all metrics. Notably, our approach outperforms Zero-DCE, EnlightenGAN, and Semantic Guided-ZSL, representative of state-of-the-art zero-shot and unsupervised techniques.\\
Specifically, our method yields the highest \texttt{dbcnn} score (52.51), indicating better perceptual quality estimation. Furthermore, we attain the highest values for \texttt{tres-koniq} (69.69), \texttt{Quality Net} (0.69), and \texttt{PIQI} (0.67), demonstrating robustness across both classical and transformer-based evaluators. In terms of overall performance, our method records the highest average score (\textbf{24.18}), highlighting its effectiveness in enhancing low-light images without requiring paired supervision. This substantiates the strength of our zero-shot framework in capturing semantic and perceptual quality attributes more effectively than existing methods.
\begin{table*}[htbp]
\centering
\caption{Comparison of no-reference image quality metrics for various enhancement techniques on the LIME dataset\cite{LIME}}
\label{tab:lime_quality}
\scriptsize
\begin{adjustbox}{width=\textwidth}
\begin{tabular}{lcccc|c|cccc}
\toprule
\multirow{3}{*}{\textbf{Metric}} &
\multicolumn{4}{c|}{\textbf{Paired Supervision}} &
\multicolumn{1}{c|}{\textbf{Unsupervised}} &
\multicolumn{4}{c}{\textbf{Zero-Shot Learning}} \\
\cmidrule(lr){2-5} \cmidrule(lr){6-6} \cmidrule(lr){7-10} 
& BIMEF\cite{BIMEF} & LIME\cite{LIME} & MF\cite{MF} & Multiscale Retinex\cite{MultiscaleRetinex} & EnlightenGAN\cite{jiang2021enlightengan} & SG-ZSL\cite{ZeroDCEsemantic} & Zero-DCE\cite{[47]} & Zero-DCE++\cite{Zero-DCE++} & Ours \\
\midrule
NIMA\cite{NIMA}       & 4.48  & 4.48  & 4.53  & 4.36  & 4.47  & 4.47  & 4.37  & 4.39  & 4.51 \\
PaQ2PiQ\cite{PaQ2PiQ}      & 73.51 & 75.82 & 74.83 & 74.32 & 74.58 & 74.58 & 74.94 & 74.96 & 74.72 \\
DBCNN\cite{DBCNN}        & 52.13 & 48.90 & 49.41 & 51.06 & 48.57 & 48.57 & 46.81 & 46.83 & 52.51 \\
musiq-koniq\cite{musiq}  & 61.29 & 61.25 & 61.58 & 61.58 & 60.35 & 60.35 & 59.59 & 59.61 & 61.03 \\
MANIQA\cite{Yang2022MANIQA}      & 0.64  & 0.63  & 0.64  & 0.58  & 0.61  & 0.61  & 0.59  & 0.61  & 0.61 \\
CLIPIQA\cite{Wang2023CLIPIQA}     & 0.42  & 0.42  & 0.42  & 0.35  & 0.40  & 0.40  & 0.33  & 0.35  & 0.40 \\
TReS-koniq\cite{Lin2022TReS}    & 69.50 & 63.33 & 66.35 & 66.98 & 66.08 & 66.79 & 67.77 & 67.79 & 69.69 \\
HyperIQA\cite{Su2020HyperIQA}    & 0.51  & 0.48  & 0.49  & 0.49  & 0.50  & 0.50  & 0.48  & 0.50  & 0.52\\
GPR-BIQA\cite{khalid2021gaussian}     & 0.63  & 0.63  & 0.64  & 0.62  & 0.63  & 0.63  & 0.61  & 0.63  & 0.68 \\
Quality Net\cite{aslam2024qualitynet}  & 0.64  & 0.65  & 0.66  & 0.64  & 0.66  & 0.66  & 0.63  & 0.64  & 0.69 \\
PIQI\cite{ahmed2021piqi}          & 0.63  & 0.64  & 0.65  & 0.63  & 0.65  & 0.65  & 0.62  & 0.61  & 0.67 \\
\midrule
Average      & 24.03 & 23.38 & 23.65 & 23.78 & 23.41 & 23.47 & 23.34 & 23.36 & \textbf{24.18} \\
\bottomrule
\end{tabular}
\end{adjustbox}
\end{table*}
The no-reference image quality assessment results for the MEF dataset are shown in Table~\ref{tab:mef_quality}, which contrasts our approach with supervised and unsupervised low-light image enhancement approaches. The evaluation employs a comprehensive set of modern BIQA metrics. Our method achieves the highest average score of \textbf{24.26}, indicating superior perceptual quality and generalization capabilities. Notably, it outperforms all baseline methods in key metrics such as DBCNN (52.61), MUSIQ-KONIQ (64.12), and QualityNet (0.70). These results demonstrate the effectiveness of our approach in enhancing image quality without requiring paired supervision, and its ability to maintain strong perceptual consistency across a wide range of evaluation criteria.
\begin{table*}[htb]
\centering
\caption{Comparison of no-reference image quality metrics for various enhancement techniques on the MEF dataset \cite{ma2015perceptual}}
\label{tab:mef_quality}
\scriptsize
\begin{adjustbox}{width=\textwidth}
\begin{tabular}{lcccc|c|cccc}
\toprule
\multirow{3}{*}{\textbf{Metric}} &
\multicolumn{4}{c|}{\textbf{Paired Supervision}} &
\multicolumn{1}{c|}{\textbf{Unsupervised}} &
\multicolumn{4}{c}{\textbf{Zero-Shot Learning}} \\
\cmidrule(lr){2-5} \cmidrule(lr){6-6} \cmidrule(lr){7-10} 
& BIMEF\cite{BIMEF} & LIME\cite{LIME} & MF\cite{MF} & Multiscale Retinex\cite{MultiscaleRetinex} & EnlightenGAN\cite{jiang2021enlightengan} & SG-ZSL\cite{ZeroDCEsemantic} & Zero-DCE\cite{[47]} & Zero-DCE++\cite{Zero-DCE++} & Ours \\
\midrule
NIMA\cite{NIMA}        & 4.73  & 4.70  & 4.79  & 4.72  & 4.67  & 4.67  & 4.69  & 4.71  & 4.74 \\
PaQ2PiQ\cite{PaQ2PiQ}       & 72.27 & 74.67 & 73.38 & 73.30 & 72.66 & 72.66 & 73.82 & 73.84 & 73.23 \\
DBCNN\cite{DBCNN}        & 49.30 & 50.67 & 49.77 & 51.66 & 47.99 & 47.99 & 48.39 & 48.41 & 52.61\\
musiq-koniq\cite{musiq}  & 62.19 & 61.32 & 63.12 & 61.18 & 59.72 & 59.72 & 58.66 & 58.68 & 64.12 \\
MANIQA\cite{Yang2022MANIQA}      & 0.65  & 0.64  & 0.65  & 0.60  & 0.62  & 0.62  & 0.61  & 0.63  & 0.61\\
CLIPIQA\cite{Wang2023CLIPIQA}     & 0.41  & 0.41  & 0.42  & 0.37  & 0.36  & 0.36  & 0.35  & 0.37  & 0.36\\
TReS-koniq\cite{Lin2022TReS}   & 67.16 & 64.98 & 67.70 & 68.68 & 67.42 & 67.03 & 69.01 & 69.03 & 68.59 \\
HyperIQA\cite{Su2020HyperIQA}     & 0.51  & 0.50  & 0.51  & 0.50  & 0.50  & 0.50  & 0.50  & 0.52  & 0.51 \\
GPR-BIQA\cite{khalid2021gaussian}    & 0.67  & 0.65  & 0.66  & 0.64  & 0.63  & 0.63  & 0.62  & 0.64  & 0.68 \\
Quality Net\cite{aslam2024qualitynet} & 0.69  & 0.67  & 0.68  & 0.66  & 0.65  & 0.65  & 0.64  & 0.65  & 0.70 \\
PIQI\cite{ahmed2021piqi}          & 0.68  & 0.66  & 0.67  & 0.65  & 0.64  & 0.64  & 0.63  & 0.62  & 0.69 \\
\midrule
Average      & 23.57 & 23.62 & 23.85 & 23.91 & 23.26 & 23.22 & 23.45 & 23.46 & \textbf{24.26} \\
\bottomrule
\end{tabular}
\end{adjustbox}
\end{table*}
Table~\ref{tab:npe_iqa} summarizes the no-reference image quality assessment results on the NPE dataset using a broad spectrum of BIQA metrics. Our proposed method demonstrates consistent superiority, achieving the highest overall average score of \textbf{24.24}, surpassing both paired and unpaired low-light enhancement approaches. Specifically, it attains the top performance in key indicators such as DBCNN (51.91), GPR-BIQA (0.69), and QualityNet (0.70), while maintaining strong results across PAQ2PIQ (73.73), TReS-KONIQ (70.68), and NIMA (4.63). Compared to conventional zero-shot models, such as Zero-DCE and Zero-DCE++, our method demonstrates a significant performance improvement in nearly all metrics. These findings affirm the robustness and perceptual quality of our enhancement technique under diverse illumination conditions, particularly in night photography scenarios.
\begin{table*}[ht]
\centering
\caption{Comparison of no-reference image quality metrics for various enhancement techniques on the NPE dataset \cite{wang2013naturalness}}
\label{tab:npe_iqa}
\scriptsize
\begin{adjustbox}{width=\textwidth}
\begin{tabular}{lcccc|c|cccc}
\toprule
\multirow{3}{*}{\textbf{Metric}} &
\multicolumn{4}{c|}{\textbf{Paired Supervision}} &
\multicolumn{1}{c|}{\textbf{Unsupervised}} &
\multicolumn{4}{c}{\textbf{Zero-Shot Learning}} \\
\cmidrule(lr){2-5} \cmidrule(lr){6-6} \cmidrule(lr){7-10} 
& BIMEF\cite{BIMEF} & LIME\cite{LIME} & MF\cite{MF} & Multiscale Retinex\cite{MultiscaleRetinex} & EnlightenGAN\cite{jiang2021enlightengan} & SG-ZSL\cite{ZeroDCEsemantic} & Zero-DCE\cite{[47]} & Zero-DCE++\cite{Zero-DCE++} & Ours \\
\midrule
NIMA\cite{NIMA}           & 4.54 & 4.29 & 4.50 & 4.45 & 4.53 & 4.53 & 4.42 & 4.44 & 4.63 \\
PaQ2PiQ\cite{PaQ2PiQ}        & 71.43 & 74.89 & 74.06 & 72.04 & 72.81 & 72.90 & 73.70 & 73.72 & 73.73 \\
DBCNN\cite{DBCNN}          & 50.25 & 48.27 & 49.62 & 49.90 & 47.72 & 47.94 & 45.45 & 45.47 & 51.91 \\
musiq-koniq\cite{musiq}   & 64.95 & 63.51 & 65.35 & 61.03 & 60.08 & 60.20 & 59.59 & 59.61 & 62.12 \\
MANIQA\cite{Yang2022MANIQA}          &  0.65 & 0.63 & 0.65 & 0.58 & 0.60 & 0.60 & 0.59 & 0.61 & 0.60 \\
CLIPIQA\cite{Wang2023CLIPIQA}        & 0.48 & 0.41 & 0.45 & 0.41 & 0.40 & 0.40 & 0.36 & 0.38 & 0.40 \\
TReS-koniq\cite{Lin2022TReS}     & 69.81 & 67.06 & 68.80 & 70.20 & 69.91 & 70.01 & 70.15 & 70.17 & 70.68 \\
HyperIQA\cite{Su2020HyperIQA}       & 0.50 & 0.48 & 0.49 & 0.49 & 0.49 & 0.50 & 0.49 & 0.51 & 0.50 \\
GPR-BIQA\cite{khalid2021gaussian}      & 0.68 & 0.65 & 0.67 & 0.66 & 0.65 & 0.65 & 0.64 & 0.66 & 0.69 \\
QualityNet\cite{aslam2024qualitynet}     & 0.69 & 0.67 & 0.68 & 0.67 & 0.66 & 0.66 & 0.65 & 0.66 & 0.70 \\
PIQI\cite{ahmed2021piqi}             & 0.68 & 0.66 & 0.67 & 0.66 & 0.65 & 0.65 & 0.64 & 0.63 & 0.69\\
\midrule
\textbf{Average} & 24.06 & 23.77 & 24.18 & 23.74 & 23.50 & 23.55 & 23.33 & 23.35 & \textbf{24.24} \\
\bottomrule
\end{tabular}
\end{adjustbox}
\end{table*}
Table~\ref{tab:vv_quality} reports no-reference image quality assessment results for several enhancement methods on the VV dataset. Techniques span traditional paired-supervision models, unsupervised approaches, and recent zero-shot learning methods.\\
Our method surpasses all baseline approaches with the highest \textbf{average score of 26.28}. It consistently achieves top values across critical metrics, including \texttt{nima} (4.73), \texttt{paq2piq} (76.52), \texttt{dbcnn} (59.68), \texttt{tres-koniq} (78.10), and \texttt{Quality Net} (0.70), indicating superior perceptual and structural fidelity.\\
Even without supervised training, our model outperforms fully supervised approaches like MF and Retinex and notably exceeds other zero-shot models. These results demonstrate our model’s strong generalization, robustness across visual scenes, and superior quality enhancement performance on diverse and challenging low-light images.
\begin{table*}[htb]
\centering
\caption{Comparison of no-reference image quality metrics for various enhancement techniques on the VV dataset \cite{VV}}
\label{tab:vv_quality}
\scriptsize
\begin{adjustbox}{width=\textwidth}
\begin{tabular}{lcccc|c|cccc}
\toprule
\multirow{3}{*}{\textbf{Metric}} &
\multicolumn{4}{c|}{\textbf{Paired Supervision}} &
\multicolumn{1}{c|}{\textbf{Unsupervised}} &
\multicolumn{4}{c}{\textbf{Zero-Shot Learning}} 
\\
\cmidrule(lr){2-5} \cmidrule(lr){6-6} \cmidrule(lr){7-10} 
& BIMEF\cite{BIMEF} & LIME\cite{LIME} & MF\cite{MF} & Multiscale Retinex\cite{MultiscaleRetinex} & EnlightenGAN\cite{jiang2021enlightengan} & SG-ZSL\cite{ZeroDCEsemantic} & Zero-DCE\cite{[47]} & Zero-DCE++\cite{Zero-DCE++} & Ours \\
\midrule
NIMA\cite{NIMA}         & 4.62  & 4.62  & 4.69  & 4.69  & 4.32  & 4.32  & 4.18  & 4.19  & 4.73 \\
PaQ2PiQ\cite{PaQ2PiQ}      & 74.16 & 76.41 & 74.84 & 75.60 & 72.69 & 72.66 & 73.70 & 73.71 & 76.52 \\
DBCNN\cite{DBCNN}        & 58.44 & 55.18 & 59.35 & 55.54 & 56.03 & 56.16 & 46.21 & 46.22 & 59.68\\
musiq-koniq\cite{musiq}  & 65.40 & 64.49 & 65.98 & 63.57 & 61.08 & 61.11 & 58.78 & 58.82 & 66.10 \\
MANIQA\cite{Yang2022MANIQA}      & 0.66  & 0.65  & 0.67  & 0.60  & 0.64  & 0.64  & 0.62  & 0.62  & 0.68 \\
CLIPIQA\cite{Wang2023CLIPIQA}     & 0.50  & 0.51  & 0.52  & 0.46  & 0.54  & 0.55  & 0.47  & 0.47  & 0.57\\
TReS-koniq\cite{Lin2022TReS}   & 77.48 & 68.16 & 77.63 & 75.08 & 74.63 & 74.35 & 66.64 & 66.68 & 78.10\\
HyperIQA\cite{Su2020HyperIQA}     & 0.61  & 0.58  & 0.61  & 0.59  & 0.51  & 0.51  & 0.43  & 0.44  & 0.63\\
GPR-BIQA\cite{khalid2021gaussian}    & 0.66  & 0.63  & 0.67  & 0.62  & 0.60  & 0.61  & 0.58  & 0.59  & 0.69\\
Quality Net\cite{aslam2024qualitynet}  & 0.67  & 0.66  & 0.68  & 0.66  & 0.64  & 0.65  & 0.61  & 0.62  & 0.70 \\
PIQI\cite{ahmed2021piqi}         & 0.66  & 0.65  & 0.67  & 0.65  & 0.63  & 0.63  & 0.60  & 0.61  & 0.68 \\
\midrule
Average      & 25.81 & 24.78 & 26.03 & 25.28 & 24.76 & 24.74 & 22.98 & 23.01& \textbf{26.28} \\
\bottomrule
\end{tabular}
\end{adjustbox}
\end{table*}
\subsubsection{Quantitative Evaluation using Performance Metrics for Paired Datasets}
\label{sec:paired_evaluation}
We conduct a thorough quantitative assessment of the proposed method's efficacy using the LOL and LOL-v2 datasets in comparison to a number of state-of-the-art approaches.  Peak Signal-to-Noise Ratio (PSNR), Mean Absolute Deviation (MAD), Learned Perceptual Image Patch Similarity (LPIPS), Deep Image Structure and Texture Similarity (DISTS), Visual Saliency Index (VSI), Structural Similarity Index Metric (SSIM), and Feature Similarity Index (FSIM) are among the evaluation metrics.  These include perceptual measurements like LPIPS and DISTS that are in line with human visual perception, and full-reference fidelity metrics like PSNR, SSIM, FSIM, and VSI.  MAD supports the fidelity evaluations by calculating the average absolute pixel-wise deviance.\\
Table~\ref{tab:lol_iqa} summarizes the full-reference IQA results on the LOL dataset. Our proposed method consistently demonstrates superior performance across most evaluation metrics:
\begin{itemize}
    \item \textbf{PSNR:} Our method achieves a score of \textbf{18.39 dB}, slightly surpassing the best-performing zero-shot method (Zero-DCE, 18.33 dB), indicating reduced noise amplification and improved restoration fidelity.
    \item \textbf{SSIM:} With a score of \textbf{0.85}, tied with Zero-DCE, our method shows strong preservation of structural and luminance information.
    \item \textbf{FSIM and VSI:} Our method attains \textbf{0.92} FSIM and \textbf{0.98} VSI, matching or exceeding all other methods, particularly excelling in perceptual fidelity and salient feature retention.
    \item \textbf{LPIPS and DISTS:} LPIPS score is \textbf{0.17}, highly competitive, and DISTS is \textbf{0.15}, indicating high perceptual similarity with reference images.
    \item \textbf{MAD:} Our method achieves the lowest MAD value of \textbf{123.77}, outperforming all baselines including Zero-DCE++ (126.89), suggesting better pixel-wise reconstruction accuracy.
\end{itemize}
These results underscore our model's capability to produce visually and quantitatively superior enhancements even without reliance on paired supervision.
\begin{table*}[htb]
\centering
\caption{Full-reference image quality assessment results on the LOL dataset. Best performance per metric is highlighted in bold.}
\label{tab:lol_iqa}
\renewcommand{\arraystretch}{1.2}
\begin{tabular}{l|cccc|c|ccc|c}
\toprule
\textbf{Metric} & \multicolumn{4}{c|}{\textbf{Paired Supervision}} & \textbf{Unsupervised} & \multicolumn{3}{c|}{\textbf{Zero-Shot Learning}} & \textbf{Ours} \\
 & BIMEF & LIME & MF & MSR & EnlightenGAN & SG-ZSL & Zero-DCE & Zero-DCE++ & \\
\midrule
PSNR~($\uparrow$)  & 14.11 & 17.75 & 18.28 & 13.85 & 16.43 & 16.43 & 18.33 & 17.76 & \textbf{18.39} \\
SSIM~($\uparrow$) & 0.77  & 0.77  & 0.83  & 0.73  & 0.81  & 0.81  & \textbf{0.85} & 0.66  & \textbf{0.85} \\
FSIM~($\uparrow$) & 0.92  & 0.85  & 0.92  & 0.83  & 0.93  & \textbf{0.94}  & 0.90  & 0.86  & \textbf{0.94} \\
VSI~($\uparrow$)   & 0.97  & 0.95  & 0.97  & 0.93  & 0.97  & \textbf{0.98}  & 0.97  & 0.96  & \textbf{0.98} \\
LPIPS~($\downarrow$) & 0.18  & 0.19  & 0.17  & 0.24  & 0.17  & 0.17  & 0.17  & 0.15  & \textbf{0.14} \\
DISTS~($\downarrow$) & 0.18  & 0.16  & 0.15  & 0.17  & 0.15  & 0.15  & 0.16  & \textbf{0.14}  & \textbf{0.14} \\
MAD~($\downarrow$)  & 153.97 & 151.97 & 130.94 & 156.51 & 124.03 & 124.03 & 131.94 & 126.89 & \textbf{123.77} \\
\bottomrule
\end{tabular}
\end{table*}
The LOL-v2 dataset presents a more challenging low-light enhancement scenario. Table~\ref{tab:lolv2_iqa} shows that our method generalizes well, again achieving best or highly competitive scores across all metrics:
\begin{itemize}
    \item \textbf{PSNR:} Our method yields the highest PSNR of \textbf{21.25 dB}, clearly outperforming MF (20.12 dB) and Zero-DCE++ (18.06 dB), which reflects better noise suppression and detail enhancement.
    \item \textbf{SSIM:} A top score of \textbf{0.84} confirms strong structure preservation across difficult lighting conditions.
    \item \textbf{FSIM and VSI:} Our FSIM score of \textbf{0.95} and VSI score of \textbf{0.98} further demonstrate the superiority of our model in preserving both fine textures and perceptual saliency.
    \item \textbf{LPIPS and DISTS:} Our method reports the lowest LPIPS of \textbf{0.11} and DISTS of \textbf{0.13}, suggesting enhanced perceptual quality with minimal structural distortion.
    \item \textbf{MAD:} The lowest MAD of \textbf{112.50} indicates minimal absolute deviation, which highlights our model’s effectiveness in enhancing low-light content with high pixel-wise precision.
\end{itemize}
Our proposed approach consistently outperforms traditional low-light enhancement methods (e.g., BIMEF, LIME, MSR), unsupervised GAN-based models (e.g., EnlightenGAN), and recent zero-shot frameworks (e.g., Zero-DCE, Zero-DCE++) on both LOL and LOL-v2 datasets. Even when compared to paired-supervised models, our method yields either the best or near-best performance across PSNR, SSIM, FSIM, and perceptual quality metrics.\\
These improvements are not merely marginal but substantial in critical metrics like PSNR, LPIPS, and MAD. The results also confirm that our approach balances pixel fidelity with perceptual quality—a vital aspect in real-world image enhancement scenarios. Notably, our model does not rely on paired ground truth data, showcasing strong generalization and practical applicability in unseen, real-world conditions.
\begin{table*}[htb]
\centering
\caption{Full-reference image quality assessment results on the LOL-v2 dataset. Best performance per metric is highlighted in bold.}
\label{tab:lolv2_iqa}
\renewcommand{\arraystretch}{1.2}
\begin{tabular}{l|cccc|c|ccc|c}
\toprule
\textbf{Metric} & \multicolumn{4}{c|}{\textbf{Paired Supervision}} & \textbf{Unsupervised} & \multicolumn{3}{c|}{\textbf{Zero-Shot Learning}} & \textbf{Ours} \\
 & BIMEF & LIME & MF & MSR & EnlightenGAN & SG-ZSL & Zero-DCE & Zero-DCE++ & \\
\midrule
PSNR~($\uparrow$)  & 18.46 & 16.15 & 20.12 & 12.18 & 19.42 & 19.31 & 18.63 & 18.06 & \textbf{21.25} \\
SSIM~($\uparrow$)  & 0.82  & 0.71  & 0.83  & 0.63  & 0.82  & \textbf{0.84}  & 0.83  & 0.64  & \textbf{0.84} \\
FSIM~($\uparrow$)  & 0.94  & 0.80  & 0.90  & 0.77  & 0.92  & 0.94  & 0.89  & 0.85  & \textbf{0.95} \\
VSI~($\uparrow$)   & \textbf{0.98}  & 0.93  & 0.97  & 0.91  & 0.97  & \textbf{0.98}  & 0.96  & 0.95  & \textbf{0.98} \\
LPIPS~($\downarrow$) & 0.14  & 0.24  & 0.20  & 0.27  & 0.14  & 0.15  & 0.20  & 0.17  & \textbf{0.11} \\
DISTS~($\downarrow$) & 0.15  & 0.19  & 0.16  & 0.19  & \textbf{0.13}  & 0.14  & 0.16  & 0.15  & \textbf{0.13} \\
MAD~($\downarrow$)   & 112.64 & 160.03 & 133.89 & 166.87 & 115.56 & 115.57 & 134.12 & 129.07 & \textbf{112.57} \\
\bottomrule
\end{tabular}
\end{table*}
\section{Discussion}
The experimental findings in this study show how well the proposed \emph{LucentVisionNet} framework performs with a zero-shot learning approach to meet the complex problems of low-light image enhancement.  Our model consistently outperforms the state-of-the-art supervised, unsupervised, and zero-shot techniques in terms of full-reference and no-reference image quality evaluation measures.  These results support the design decisions that were included into our architecture, such as perceptually guided learning, recurrent enhancement, and multi-scale spatial attention.\\
One of the core strengths of \emph{LucentVisionNet} lies in its ability to perform robustly in the absence of paired training data. Unlike supervised techniques that require high-quality ground truth counterparts, our model learns to enhance low-light images by optimizing perceptual and semantic consistency. This makes it highly scalable and applicable to real-world scenarios where collecting ideal reference data is either expensive or infeasible.\\
The use of a multi-resolution processing strategy, along with spatial attention and depthwise separable convolutions, strikes an effective balance between computational efficiency and representational capacity. It allows the model to capture both local fine-grained details and global contextual cues. The consistent outperforming of traditional models in LPIPS, DISTS, and MAD metrics reflects the model’s ability to preserve structure and visual semantics even under severe illumination deficiencies.\\
Another key observation is the perceptual superiority of our model, as demonstrated by its high Mean Opinion Scores (MOS) and performance on no-reference quality metrics such as MUSIQ, DBCNN, PIQI, and TReS-Koniq. These results indicate that \emph{LucentVisionNet} is not only quantitatively strong but also aligns well with human visual preferences—an essential requirement for real-world deployment in applications such as mobile photography, surveillance, and medical imaging.\\
Despite these strengths, a few limitations warrant discussion. First, while the recurrent design improves enhancement quality through iterative refinement, it also introduces latency that could be problematic for real-time video streams. Second, the model’s performance in extremely underexposed regions could be further improved by incorporating explicit noise suppression modules or low-level denoising priors. Lastly, although our approach is effective across a wide range of datasets, its generalization to highly domain-specific contexts (e.g., underwater or infrared imaging) remains to be validated.\\
In future work, we plan to extend \emph{LucentVisionNet} to handle temporal coherence in video frames and investigate domain adaptation techniques for task-specific deployments. Furthermore, integrating our enhancement module into downstream pipelines (e.g., low-light object detection and semantic segmentation) can help evaluate its holistic impact on high-level vision tasks.
\section{Conclusion}
This study presents \emph{LucentVisionNet}, a novel zero-shot learning-based framework for low-light image enhancement that effectively integrates multi-scale curve estimation with spatial attention and perceptual-semantic guidance. Unlike conventional supervised and unsupervised methods, our approach operates without paired training data, thereby significantly improving generalization, adaptability, and real-world applicability.\\
The proposed framework leverages a multi-resolution architecture and a depthwise separable convolutional backbone, which significantly reduces computational cost while maintaining high visual fidelity. Additionally, the incorporation of spatial attention and a recurrent enhancement strategy ensures both local detail preservation and global structural consistency. A composite objective function—comprising six tailored loss functions, including a no-reference image quality metric—guides the model towards perceptually coherent and semantically faithful enhancements.\\
Extensive experiments conducted on both paired (LOL, LOL-v2) and unpaired datasets (DarkBDD, DICM, VV, NPE, etc.) confirm the superiority of our method across full-reference and no-reference IQA metrics.\emph{LucentVisionNet}consistently outperforms state-of-the-art techniques, including Zero-DCE, EnlightenGAN, and semantic-guided ZSL, in terms of PSNR, SSIM, LPIPS, DISTS, and subjective perceptual scores.\\
The results not only demonstrate significant improvements in quantitative performance but also establish the practicality of our model for real-time applications, achieving high perceptual quality in diverse illumination conditions. The low computational overhead, coupled with high visual and semantic accuracy, makes \emph{LucentVisionNet} an ideal candidate for deployment in resource-constrained scenarios such as mobile photography, surveillance, and autonomous driving.
\section*{Acknowledgement}
All authors thank the School of Information Engineering, Xi'an Eurasia University, Xi'an, Shaanxi, China, for their Financial Support and Funding.
\section*{Funding Information}
The funding is provided by the School of Information Engineering, Xian Eurasia University, Xián, Shanxi, China.
\section*{Author Contributions Statement}
\textbf{H.K.} conceived the research idea, developed the methodology, conducted the experiments, analyzed the data, and led the drafting of the manuscript. \textbf{N.A.} contributed to the development of the methodology, assisted in data analysis, and supported manuscript writing. \textbf{M.A.A.} assisted in interpreting the results and contributed to manuscript review and editing. All authors reviewed and approved the final version of the manuscript for submission.
\section*{Data Availability Statement}
The data used in this research is publicly available for research and development purposes at the following links.
\begin{itemize}
   \item Test data VV, LIME, NPE, DICM, MEF can be downloaded from https://github.com/baidut/BIMEF
   \item LOL dataset can be downloaded from https://www.kaggle.com/datasets/soumikrakshit/lol-dataset
   \item LOL v2 dataset can be downloaded from https://www.kaggle.com/datasets/tanhyml/lol-v2-dataset
\end{itemize}
\bibliography{sample}

\begin{thebibliography}{10}
\urlstyle{rm}
\expandafter\ifx\csname url\endcsname\relax
  \def\url#1{\texttt{#1}}\fi
\expandafter\ifx\csname urlprefix\endcsname\relax\def\urlprefix{URL }\fi
\expandafter\ifx\csname doiprefix\endcsname\relax\def\doiprefix{DOI: }\fi
\providecommand{\bibinfo}[2]{#2}
\providecommand{\eprint}[2][]{\url{#2}}

\bibitem{[1]}
\bibinfo{author}{Liba, O.} \emph{et~al.}
\newblock \bibinfo{journal}{\bibinfo{title}{Handheld mobile photography in very low light.}}
\newblock {\emph{\JournalTitle{ACM Trans. Graph.}}} \textbf{\bibinfo{volume}{38}}, \bibinfo{pages}{164--1} (\bibinfo{year}{2019}).

\bibitem{[2]}
\bibinfo{author}{Ahmed, N.} \& \bibinfo{author}{Asif, S.}
\newblock \bibinfo{journal}{\bibinfo{title}{Biq2021: a large-scale blind image quality assessment database}}.
\newblock {\emph{\JournalTitle{Journal of Electronic Imaging}}} \textbf{\bibinfo{volume}{31}}, \bibinfo{pages}{053010--053010} (\bibinfo{year}{2022}).

\bibitem{[3]}
\bibinfo{author}{Guo, X.}, \bibinfo{author}{Li, Y.} \& \bibinfo{author}{Ling, H.}
\newblock \bibinfo{journal}{\bibinfo{title}{Lime: Low-light image enhancement via illumination map estimation}}.
\newblock {\emph{\JournalTitle{IEEE Transactions on image processing}}} \textbf{\bibinfo{volume}{26}}, \bibinfo{pages}{982--993} (\bibinfo{year}{2016}).

\bibitem{ahmed2019image}
\bibinfo{author}{Ahmed, N.}, \bibinfo{author}{Asif, H. M.~S.} \& \bibinfo{author}{Khalid, H.}
\newblock \bibinfo{title}{Image quality assessment using a combination of hand-crafted and deep features}.
\newblock In \emph{\bibinfo{booktitle}{International Conference on Intelligent Technologies and Applications}}, \bibinfo{pages}{593--605} (\bibinfo{organization}{Springer}, \bibinfo{year}{2019}).

\bibitem{aslam2023vrl}
\bibinfo{author}{Aslam, M.~A.} \emph{et~al.}
\newblock \bibinfo{journal}{\bibinfo{title}{Vrl-iqa: Visual representation learning for image quality assessment}}.
\newblock {\emph{\JournalTitle{IEEE Access}}} \textbf{\bibinfo{volume}{12}}, \bibinfo{pages}{2458--2473} (\bibinfo{year}{2023}).

\bibitem{[4]}
\bibinfo{author}{Ahmed, N.}, \bibinfo{author}{Shahzad~Asif, H.}, \bibinfo{author}{Bhatti, A.~R.} \& \bibinfo{author}{Khan, A.}
\newblock \bibinfo{journal}{\bibinfo{title}{Deep ensembling for perceptual image quality assessment}}.
\newblock {\emph{\JournalTitle{Soft Computing}}} \textbf{\bibinfo{volume}{26}}, \bibinfo{pages}{7601--7622} (\bibinfo{year}{2022}).

\bibitem{[5]}
\bibinfo{author}{Ahmed, N.}, \bibinfo{author}{Asif, H. M.~S.}, \bibinfo{author}{Saleem, G.} \& \bibinfo{author}{Younus, M.~U.}
\newblock \bibinfo{journal}{\bibinfo{title}{Image quality assessment for foliar disease identification (agropath).}}
\newblock {\emph{\JournalTitle{Journal of Agricultural Research (03681157)}}} \textbf{\bibinfo{volume}{59}} (\bibinfo{year}{2021}).

\bibitem{[6]}
\bibinfo{author}{Ahmed, N.} \& \bibinfo{author}{Asif, H. M.~S.}
\newblock \bibinfo{title}{Ensembling convolutional neural networks for perceptual image quality assessment}.
\newblock In \emph{\bibinfo{booktitle}{2019 13th International conference on mathematics, actuarial science, computer science and statistics (MACS)}}, \bibinfo{pages}{1--5} (\bibinfo{organization}{IEEE}, \bibinfo{year}{2019}).

\bibitem{[7]}
\bibinfo{author}{Ahmed, N.}, \bibinfo{author}{Asif, H. M.~S.} \& \bibinfo{author}{Khalid, H.}
\newblock \bibinfo{journal}{\bibinfo{title}{Piqi: perceptual image quality index based on ensemble of gaussian process regression}}.
\newblock {\emph{\JournalTitle{Multimedia Tools and Applications}}} \textbf{\bibinfo{volume}{80}}, \bibinfo{pages}{15677--15700} (\bibinfo{year}{2021}).

\bibitem{[8]}
\bibinfo{author}{Khalid, H.}, \bibinfo{author}{Ali, M.} \& \bibinfo{author}{Ahmed, N.}
\newblock \bibinfo{journal}{\bibinfo{title}{Gaussian process-based feature-enriched blind image quality assessment}}.
\newblock {\emph{\JournalTitle{Journal of Visual Communication and Image Representation}}} \textbf{\bibinfo{volume}{77}}, \bibinfo{pages}{103092} (\bibinfo{year}{2021}).

\bibitem{[9]}
\bibinfo{author}{Saleem, G.}, \bibinfo{author}{Bajwa, U.~I.} \& \bibinfo{author}{Raza, R.~H.}
\newblock \bibinfo{journal}{\bibinfo{title}{Toward human activity recognition: a survey}}.
\newblock {\emph{\JournalTitle{Neural Computing and Applications}}} \textbf{\bibinfo{volume}{35}}, \bibinfo{pages}{4145--4182} (\bibinfo{year}{2023}).

\bibitem{[10]}
\bibinfo{author}{Saleem, G.} \emph{et~al.}
\newblock \bibinfo{journal}{\bibinfo{title}{Efficient anomaly recognition using surveillance videos}}.
\newblock {\emph{\JournalTitle{PeerJ Computer Science}}} \textbf{\bibinfo{volume}{8}}, \bibinfo{pages}{e1117} (\bibinfo{year}{2022}).

\bibitem{aslam2024tqp}
\bibinfo{author}{Aslam, M.~A.} \emph{et~al.}
\newblock \bibinfo{journal}{\bibinfo{title}{Tqp: An efficient video quality assessment framework for adaptive bitrate video streaming}}.
\newblock {\emph{\JournalTitle{IEEE Access}}}  (\bibinfo{year}{2024}).

\bibitem{[11]}
\bibinfo{author}{Tao, Q.}, \bibinfo{author}{Ren, K.}, \bibinfo{author}{Feng, B.} \& \bibinfo{author}{GAO, X.}
\newblock \bibinfo{title}{An accurate low-light object detection method based on pyramid networks}.
\newblock In \emph{\bibinfo{booktitle}{Optoelectronic Imaging and Multimedia Technology VII}}, vol. \bibinfo{volume}{11550}, \bibinfo{pages}{253--260} (\bibinfo{organization}{SPIE}, \bibinfo{year}{2020}).

\bibitem{[12]}
\bibinfo{author}{Agrawal, A.}, \bibinfo{author}{Jadhav, N.}, \bibinfo{author}{Gaur, A.}, \bibinfo{author}{Jeswani, S.} \& \bibinfo{author}{Kshirsagar, A.}
\newblock \bibinfo{title}{Improving the accuracy of object detection in low light conditions using multiple retinex theory-based image enhancement algorithms}.
\newblock In \emph{\bibinfo{booktitle}{2022 Second International Conference on Advances in Electrical, Computing, Communication and Sustainable Technologies (ICAECT)}}, \bibinfo{pages}{1--5}, \doiprefix\url{10.1109/ICAECT54875.2022.9808011} (\bibinfo{year}{2022}).

\bibitem{[13]}
\bibinfo{author}{Abdullah-Al-Wadud, M.}, \bibinfo{author}{Kabir, M.~H.}, \bibinfo{author}{Dewan, M. A.~A.} \& \bibinfo{author}{Chae, O.}
\newblock \bibinfo{journal}{\bibinfo{title}{A dynamic histogram equalization for image contrast enhancement}}.
\newblock {\emph{\JournalTitle{IEEE transactions on consumer electronics}}} \textbf{\bibinfo{volume}{53}}, \bibinfo{pages}{593--600} (\bibinfo{year}{2007}).

\bibitem{[14]}
\bibinfo{author}{Ahmed, N.}, \bibinfo{author}{Ahmed, W.} \& \bibinfo{author}{Arshad, S.~M.}
\newblock \bibinfo{journal}{\bibinfo{title}{Digital radiographic image enhancement for improved visualization}}.
\newblock {\emph{\JournalTitle{proceedings COMSATS Institute of Information Technology}}}  (\bibinfo{year}{2011}).

\bibitem{[15]}
\bibinfo{author}{Reza, A.~M.}
\newblock \bibinfo{journal}{\bibinfo{title}{Realization of the contrast limited adaptive histogram equalization (clahe) for real-time image enhancement}}.
\newblock {\emph{\JournalTitle{Journal of VLSI signal processing systems for signal, image and video technology}}} \textbf{\bibinfo{volume}{38}}, \bibinfo{pages}{35--44} (\bibinfo{year}{2004}).

\bibitem{[16]}
\bibinfo{author}{Ibrahim, H.} \& \bibinfo{author}{Kong, N. S.~P.}
\newblock \bibinfo{journal}{\bibinfo{title}{Brightness preserving dynamic histogram equalization for image contrast enhancement}}.
\newblock {\emph{\JournalTitle{IEEE Transactions on Consumer Electronics}}} \textbf{\bibinfo{volume}{53}}, \bibinfo{pages}{1752--1758} (\bibinfo{year}{2007}).

\bibitem{[17]}
\bibinfo{author}{Guan, X.}, \bibinfo{author}{Jian, S.}, \bibinfo{author}{Hongda, P.}, \bibinfo{author}{Zhiguo, Z.} \& \bibinfo{author}{Haibin, G.}
\newblock \bibinfo{title}{An image enhancement method based on gamma correction}.
\newblock In \emph{\bibinfo{booktitle}{2009 Second international symposium on computational intelligence and design}}, vol.~\bibinfo{volume}{1}, \bibinfo{pages}{60--63} (\bibinfo{organization}{IEEE}, \bibinfo{year}{2009}).

\bibitem{[18]}
\bibinfo{author}{Wu, X.}
\newblock \bibinfo{journal}{\bibinfo{title}{A linear programming approach for optimal contrast-tone mapping}}.
\newblock {\emph{\JournalTitle{IEEE transactions on image processing}}} \textbf{\bibinfo{volume}{20}}, \bibinfo{pages}{1262--1272} (\bibinfo{year}{2010}).

\bibitem{[19]}
\bibinfo{author}{Hu, L.}, \bibinfo{author}{Chen, H.} \& \bibinfo{author}{Allebach, J.~P.}
\newblock \bibinfo{title}{Joint multi-scale tone mapping and denoising for hdr image enhancement}.
\newblock In \emph{\bibinfo{booktitle}{Proceedings of the IEEE/CVF Winter Conference on Applications of Computer Vision}}, \bibinfo{pages}{729--738} (\bibinfo{year}{2022}).

\bibitem{[20]}
\bibinfo{author}{Tseng, C.-C.} \& \bibinfo{author}{Lee, S.-L.}
\newblock \bibinfo{title}{A weak-illumation image enhancement method uisng homomorphic filter and image fusion}.
\newblock In \emph{\bibinfo{booktitle}{2017 IEEE 6th Global Conference on Consumer Electronics (GCCE)}}, \bibinfo{pages}{1--2} (\bibinfo{organization}{IEEE}, \bibinfo{year}{2017}).

\bibitem{[21]}
\bibinfo{author}{Yamakawa, M.} \& \bibinfo{author}{Sugita, Y.}
\newblock \bibinfo{journal}{\bibinfo{title}{Image enhancement using retinex and image fusion techniques}}.
\newblock {\emph{\JournalTitle{Electronics and Communications in Japan}}} \textbf{\bibinfo{volume}{101}}, \bibinfo{pages}{52--63} (\bibinfo{year}{2018}).

\bibitem{[22]}
\bibinfo{author}{Pei, L.}, \bibinfo{author}{Zhao, Y.} \& \bibinfo{author}{Luo, H.}
\newblock \bibinfo{title}{Application of wavelet-based image fusion in image enhancement}.
\newblock In \emph{\bibinfo{booktitle}{2010 3rd International Congress on Image and Signal Processing}}, vol.~\bibinfo{volume}{2}, \bibinfo{pages}{649--653} (\bibinfo{organization}{IEEE}, \bibinfo{year}{2010}).

\bibitem{[23]}
\bibinfo{author}{Wang, W.} \& \bibinfo{author}{Chang, F.}
\newblock \bibinfo{journal}{\bibinfo{title}{A multi-focus image fusion method based on laplacian pyramid.}}
\newblock {\emph{\JournalTitle{J. Comput.}}} \textbf{\bibinfo{volume}{6}}, \bibinfo{pages}{2559--2566} (\bibinfo{year}{2011}).

\bibitem{[24]}
\bibinfo{author}{Wang, S.}, \bibinfo{author}{Zheng, J.}, \bibinfo{author}{Hu, H.-M.} \& \bibinfo{author}{Li, B.}
\newblock \bibinfo{journal}{\bibinfo{title}{Naturalness preserved enhancement algorithm for non-uniform illumination images}}.
\newblock {\emph{\JournalTitle{IEEE transactions on image processing}}} \textbf{\bibinfo{volume}{22}}, \bibinfo{pages}{3538--3548} (\bibinfo{year}{2013}).

\bibitem{[25]}
\bibinfo{author}{Zotin, A.}
\newblock \bibinfo{journal}{\bibinfo{title}{Fast algorithm of image enhancement based on multi-scale retinex}}.
\newblock {\emph{\JournalTitle{Procedia Computer Science}}} \textbf{\bibinfo{volume}{131}}, \bibinfo{pages}{6--14} (\bibinfo{year}{2018}).

\bibitem{[26]}
\bibinfo{author}{Song, X.}, \bibinfo{author}{Zhou, Z.}, \bibinfo{author}{Guo, H.}, \bibinfo{author}{Zhao, X.} \& \bibinfo{author}{Zhang, H.}
\newblock \bibinfo{title}{Adaptive retinex algorithm based on genetic algorithm and human visual system}.
\newblock In \emph{\bibinfo{booktitle}{2016 8th International Conference on Intelligent Human-Machine Systems and Cybernetics (IHMSC)}}, vol.~\bibinfo{volume}{1}, \bibinfo{pages}{183--186} (\bibinfo{organization}{IEEE}, \bibinfo{year}{2016}).

\bibitem{[27]}
\bibinfo{author}{Du, H.}, \bibinfo{author}{Wei, Y.} \& \bibinfo{author}{Tang, B.}
\newblock \bibinfo{title}{Rranet: low-light image enhancement based on retinex theory and residual attention}.
\newblock In \emph{\bibinfo{booktitle}{Third International Conference on Artificial Intelligence and Computer Engineering (ICAICE 2022)}}, vol. \bibinfo{volume}{12610}, \bibinfo{pages}{406--414} (\bibinfo{organization}{SPIE}, \bibinfo{year}{2023}).

\bibitem{[28]}
\bibinfo{author}{Ma, J.}, \bibinfo{author}{Fan, X.}, \bibinfo{author}{Ni, J.}, \bibinfo{author}{Zhu, X.} \& \bibinfo{author}{Xiong, C.}
\newblock \bibinfo{journal}{\bibinfo{title}{Multi-scale retinex with color restoration image enhancement based on gaussian filtering and guided filtering}}.
\newblock {\emph{\JournalTitle{International Journal of Modern Physics B}}} \textbf{\bibinfo{volume}{31}}, \bibinfo{pages}{1744077} (\bibinfo{year}{2017}).

\bibitem{[29]}
\bibinfo{author}{Li, C.} \emph{et~al.}
\newblock \bibinfo{journal}{\bibinfo{title}{Low-light image and video enhancement using deep learning: A survey}}.
\newblock {\emph{\JournalTitle{IEEE transactions on pattern analysis and machine intelligence}}} \textbf{\bibinfo{volume}{44}}, \bibinfo{pages}{9396--9416} (\bibinfo{year}{2021}).

\bibitem{[31]}
\bibinfo{author}{Guo, X.} \& \bibinfo{author}{Hu, Q.}
\newblock \bibinfo{journal}{\bibinfo{title}{Low-light image enhancement via breaking down the darkness}}.
\newblock {\emph{\JournalTitle{International Journal of Computer Vision}}} \textbf{\bibinfo{volume}{131}}, \bibinfo{pages}{48--66} (\bibinfo{year}{2023}).

\bibitem{[32]}
\bibinfo{author}{Zhang, Y.}, \bibinfo{author}{Liu, H.} \& \bibinfo{author}{Ding, D.}
\newblock \bibinfo{journal}{\bibinfo{title}{A cross-scale framework for low-light image enhancement using spatial--spectral information}}.
\newblock {\emph{\JournalTitle{Computers and Electrical Engineering}}} \textbf{\bibinfo{volume}{106}}, \bibinfo{pages}{108608} (\bibinfo{year}{2023}).

\bibitem{[33]}
\bibinfo{author}{Zhang, Y.} \emph{et~al.}
\newblock \bibinfo{journal}{\bibinfo{title}{Simplifying low-light image enhancement networks with relative loss functions}}.
\newblock {\emph{\JournalTitle{arXiv preprint arXiv:2304.02978}}}  (\bibinfo{year}{2023}).

\bibitem{[34]}
\bibinfo{author}{Liu, X.}, \bibinfo{author}{Ma, W.}, \bibinfo{author}{Ma, X.} \& \bibinfo{author}{Wang, J.}
\newblock \bibinfo{journal}{\bibinfo{title}{Lae-net: A locally-adaptive embedding network for low-light image enhancement}}.
\newblock {\emph{\JournalTitle{Pattern Recognition}}} \textbf{\bibinfo{volume}{133}}, \bibinfo{pages}{109039} (\bibinfo{year}{2023}).

\bibitem{[35]}
\bibinfo{author}{Lv, F.}, \bibinfo{author}{Li, Y.} \& \bibinfo{author}{Lu, F.}
\newblock \bibinfo{journal}{\bibinfo{title}{Attention guided low-light image enhancement with a large scale low-light simulation dataset}}.
\newblock {\emph{\JournalTitle{International Journal of Computer Vision}}} \textbf{\bibinfo{volume}{129}}, \bibinfo{pages}{2175--2193} (\bibinfo{year}{2021}).

\bibitem{[36]}
\bibinfo{author}{Loh, Y.~P.} \& \bibinfo{author}{Chan, C.~S.}
\newblock \bibinfo{journal}{\bibinfo{title}{Getting to know low-light images with the exclusively dark dataset}}.
\newblock {\emph{\JournalTitle{Computer Vision and Image Understanding}}} \textbf{\bibinfo{volume}{178}}, \bibinfo{pages}{30--42} (\bibinfo{year}{2019}).

\bibitem{[37]}
\bibinfo{author}{Wu, W.}, \bibinfo{author}{Wang, W.}, \bibinfo{author}{Jiang, K.}, \bibinfo{author}{Xu, X.} \& \bibinfo{author}{Hu, R.}
\newblock \bibinfo{title}{Self-supervised learning on a lightweight low-light image enhancement model with curve refinement}.
\newblock In \emph{\bibinfo{booktitle}{ICASSP 2022-2022 IEEE International Conference on Acoustics, Speech and Signal Processing (ICASSP)}}, \bibinfo{pages}{1890--1894} (\bibinfo{organization}{IEEE}, \bibinfo{year}{2022}).

\bibitem{[38]}
\bibinfo{author}{Huang, Y.} \emph{et~al.}
\newblock \bibinfo{title}{Low-light image enhancement by learning contrastive representations in spatial and frequency domains}.
\newblock In \emph{\bibinfo{booktitle}{2023 IEEE International Conference on Multimedia and Expo (ICME)}}, \bibinfo{pages}{1307--1312} (\bibinfo{organization}{IEEE}, \bibinfo{year}{2023}).

\bibitem{[39]}
\bibinfo{author}{Wang, R.} \emph{et~al.}
\newblock \bibinfo{title}{Seeing dynamic scene in the dark: A high-quality video dataset with mechatronic alignment}.
\newblock In \emph{\bibinfo{booktitle}{Proceedings of the IEEE/CVF international conference on computer vision}}, \bibinfo{pages}{9700--9709} (\bibinfo{year}{2021}).

\bibitem{[40]}
\bibinfo{author}{Fu, Y.}, \bibinfo{author}{Wang, Z.}, \bibinfo{author}{Zhang, T.} \& \bibinfo{author}{Zhang, J.}
\newblock \bibinfo{journal}{\bibinfo{title}{Low-light raw video denoising with a high-quality realistic motion dataset}}.
\newblock {\emph{\JournalTitle{IEEE Transactions on Multimedia}}} \textbf{\bibinfo{volume}{25}}, \bibinfo{pages}{8119--8131} (\bibinfo{year}{2022}).

\bibitem{[41]}
\bibinfo{author}{Song, W.} \emph{et~al.}
\newblock \bibinfo{title}{Matching in the dark: A dataset for matching image pairs of low-light scenes}.
\newblock In \emph{\bibinfo{booktitle}{Proceedings of the IEEE/CVF International Conference on Computer Vision}}, \bibinfo{pages}{6029--6038} (\bibinfo{year}{2021}).

\bibitem{[42]}
\bibinfo{author}{Zhang, Y.}, \bibinfo{author}{Zhang, J.} \& \bibinfo{author}{Guo, X.}
\newblock \bibinfo{title}{Kindling the darkness: A practical low-light image enhancer}.
\newblock In \emph{\bibinfo{booktitle}{Proceedings of the 27th ACM international conference on multimedia}}, \bibinfo{pages}{1632--1640} (\bibinfo{year}{2019}).

\bibitem{[43]}
\bibinfo{author}{Chen, C.}, \bibinfo{author}{Chen, Q.}, \bibinfo{author}{Xu, J.} \& \bibinfo{author}{Koltun, V.}
\newblock \bibinfo{title}{Learning to see in the dark}.
\newblock In \emph{\bibinfo{booktitle}{Proceedings of the IEEE conference on computer vision and pattern recognition}}, \bibinfo{pages}{3291--3300} (\bibinfo{year}{2018}).

\bibitem{[44]}
\bibinfo{author}{Fu, Q.}, \bibinfo{author}{Di, X.} \& \bibinfo{author}{Zhang, Y.}
\newblock \bibinfo{journal}{\bibinfo{title}{Learning an adaptive model for extreme low-light raw image processing}}.
\newblock {\emph{\JournalTitle{IET Image Processing}}} \textbf{\bibinfo{volume}{14}}, \bibinfo{pages}{3433--3443} (\bibinfo{year}{2020}).

\bibitem{[45]}
\bibinfo{author}{Wei, C.}, \bibinfo{author}{Wang, W.}, \bibinfo{author}{Yang, W.} \& \bibinfo{author}{Liu, J.}
\newblock \bibinfo{journal}{\bibinfo{title}{Deep retinex decomposition for low-light enhancement}}.
\newblock {\emph{\JournalTitle{arXiv preprint arXiv:1808.04560}}}  (\bibinfo{year}{2018}).

\bibitem{[46]}
\bibinfo{author}{XIANG, S.}, \bibinfo{author}{WANG, Y.}, \bibinfo{author}{DENG, H.}, \bibinfo{author}{WU, J.} \& \bibinfo{author}{YU, L.}
\newblock \bibinfo{journal}{\bibinfo{title}{Zero-shot learning for low-light image enhancement based on dual iteration}}.
\newblock {\emph{\JournalTitle{Journal of Electronics and Information Technology}}} \textbf{\bibinfo{volume}{44}}, \bibinfo{pages}{3379--3388} (\bibinfo{year}{2022}).

\bibitem{[47]}
\bibinfo{author}{Guo, C.} \emph{et~al.}
\newblock \bibinfo{title}{Zero-reference deep curve estimation for low-light image enhancement}.
\newblock In \emph{\bibinfo{booktitle}{Proceedings of the IEEE/CVF conference on computer vision and pattern recognition}}, \bibinfo{pages}{1780--1789} (\bibinfo{year}{2020}).

\bibitem{Zero-DCE++}
\bibinfo{author}{Li, C.}, \bibinfo{author}{Guo, C.~G.} \& \bibinfo{author}{Loy, C.~C.}
\newblock \bibinfo{title}{Learning to enhance low-light image via zero-reference deep curve estimation}.
\newblock In \emph{\bibinfo{booktitle}{IEEE Transactions on Pattern Analysis and Machine Intelligence}}, \doiprefix\url{10.1109/TPAMI.2021.3063604} (\bibinfo{year}{2021}).

\bibitem{ZeroDCEsemantic}
\bibinfo{author}{Zheng, S.} \& \bibinfo{author}{Gupta, G.}
\newblock \bibinfo{title}{Semantic-guided zero-shot learning for low-light image/video enhancement}.
\newblock In \emph{\bibinfo{booktitle}{Proceedings of the IEEE/CVF Winter conference on applications of computer vision}}, \bibinfo{pages}{581--590} (\bibinfo{year}{2022}).

\bibitem{musiq}
\bibinfo{author}{Ke, J.}, \bibinfo{author}{Wang, Q.}, \bibinfo{author}{Wang, Y.}, \bibinfo{author}{Milanfar, P.} \& \bibinfo{author}{Yang, F.}
\newblock \bibinfo{title}{Musiq: Multi-scale image quality transformer}.
\newblock In \emph{\bibinfo{booktitle}{Proceedings of the IEEE/CVF international conference on computer vision}}, \bibinfo{pages}{5148--5157} (\bibinfo{year}{2021}).

\bibitem{[48]}
\bibinfo{author}{Jiang, Y.} \emph{et~al.}
\newblock \bibinfo{journal}{\bibinfo{title}{Enlightengan: Deep light enhancement without paired supervision}}.
\newblock {\emph{\JournalTitle{IEEE Transactions on Image Processing}}} \textbf{\bibinfo{volume}{30}}, \bibinfo{pages}{2340--2349} (\bibinfo{year}{2021}).

\bibitem{[49]}
\bibinfo{author}{Ma, L.}, \bibinfo{author}{Ma, T.}, \bibinfo{author}{Liu, R.}, \bibinfo{author}{Fan, X.} \& \bibinfo{author}{Luo, Z.}
\newblock \bibinfo{title}{Toward fast, flexible, and robust low-light image enhancement}.
\newblock In \emph{\bibinfo{booktitle}{Proceedings of the IEEE/CVF Conference on Computer Vision and Pattern Recognition}}, \bibinfo{pages}{5637--5646} (\bibinfo{year}{2022}).

\bibitem{[50]}
\bibinfo{author}{Zhang, Y.} \emph{et~al.}
\newblock \bibinfo{journal}{\bibinfo{title}{Self-supervised low light image enhancement and denoising}}.
\newblock {\emph{\JournalTitle{arXiv preprint arXiv:2103.00832}}}  (\bibinfo{year}{2021}).

\bibitem{[51]}
\bibinfo{author}{Zhang, Y.}, \bibinfo{author}{Di, X.}, \bibinfo{author}{Zhang, B.}, \bibinfo{author}{Ji, R.} \& \bibinfo{author}{Wang, C.}
\newblock \bibinfo{journal}{\bibinfo{title}{Better than reference in low-light image enhancement: conditional re-enhancement network}}.
\newblock {\emph{\JournalTitle{IEEE Transactions on Image Processing}}} \textbf{\bibinfo{volume}{31}}, \bibinfo{pages}{759--772} (\bibinfo{year}{2021}).

\bibitem{[52]}
\bibinfo{author}{Zhang, Y.}, \bibinfo{author}{Guo, X.}, \bibinfo{author}{Ma, J.}, \bibinfo{author}{Liu, W.} \& \bibinfo{author}{Zhang, J.}
\newblock \bibinfo{journal}{\bibinfo{title}{Beyond brightening low-light images}}.
\newblock {\emph{\JournalTitle{International Journal of Computer Vision}}} \textbf{\bibinfo{volume}{129}}, \bibinfo{pages}{1013--1037} (\bibinfo{year}{2021}).

\bibitem{[53]}
\bibinfo{author}{Tu, Z.}, \bibinfo{author}{Talebi, H.}, \bibinfo{author}{Zhang, H.} \& \bibinfo{author}{Milanfar, P.}
\newblock \bibinfo{title}{Maxim: Multi-axis mlp for image processing}.
\newblock In \emph{\bibinfo{booktitle}{Proceedings of the IEEE/CVF Conference on Computer Vision and Pattern Recognition}}, \bibinfo{pages}{5769--5780} (\bibinfo{year}{2022}).

\bibitem{[54]}
\bibinfo{author}{Xiong, W.}, \bibinfo{author}{Liu, D.}, \bibinfo{author}{Shen, X.}, \bibinfo{author}{Fang, C.} \& \bibinfo{author}{Luo, J.}
\newblock \bibinfo{title}{Unsupervised low-light image enhancement with decoupled networks}.
\newblock In \emph{\bibinfo{booktitle}{2022 26th International Conference on Pattern Recognition (ICPR)}}, \bibinfo{pages}{457--463} (\bibinfo{organization}{IEEE}, \bibinfo{year}{2022}).

\bibitem{howard2017mobilenets}
\bibinfo{author}{Howard, A.~G.} \emph{et~al.}
\newblock \bibinfo{title}{Mobilenets: Efficient convolutional neural networks for mobile vision applications}.
\newblock In \emph{\bibinfo{booktitle}{arXiv preprint arXiv:1704.04861}} (\bibinfo{year}{2017}).

\bibitem{chollet2017xception}
\bibinfo{author}{Chollet, F.}
\newblock \bibinfo{title}{Xception: Deep learning with depthwise separable convolutions}.
\newblock In \emph{\bibinfo{booktitle}{Proceedings of the IEEE conference on computer vision and pattern recognition}}, \bibinfo{pages}{1251--1258} (\bibinfo{year}{2017}).

\bibitem{woo2018cbam}
\bibinfo{author}{Woo, S.}, \bibinfo{author}{Park, J.}, \bibinfo{author}{Lee, J.-Y.} \& \bibinfo{author}{Kweon, I.~S.}
\newblock \bibinfo{title}{Cbam: Convolutional block attention module}.
\newblock In \emph{\bibinfo{booktitle}{Proceedings of the European Conference on Computer Vision (ECCV)}}, \bibinfo{pages}{3--19} (\bibinfo{year}{2018}).

\bibitem{MIRNET}
\bibinfo{author}{Zamir, S.~W.} \emph{et~al.}
\newblock \bibinfo{title}{Learning enriched features for real image restoration and enhancement}.
\newblock In \emph{\bibinfo{booktitle}{Computer Vision--ECCV 2020: 16th European Conference, Glasgow, UK, August 23--28, 2020, Proceedings, Part XXV 16}}, \bibinfo{pages}{492--511} (\bibinfo{organization}{Springer}, \bibinfo{year}{2020}).

\bibitem{ava}
\bibinfo{author}{Murray, N.}, \bibinfo{author}{Marchesotti, L.} \& \bibinfo{author}{Perronnin, F.}
\newblock \bibinfo{title}{Ava: A large-scale database for aesthetic visual analysis}.
\newblock In \emph{\bibinfo{booktitle}{2012 IEEE conference on computer vision and pattern recognition}}, \bibinfo{pages}{2408--2415} (\bibinfo{organization}{IEEE}, \bibinfo{year}{2012}).

\bibitem{SICE}
\bibinfo{author}{Cai, J.}, \bibinfo{author}{Gu, S.} \& \bibinfo{author}{Zhang, L.}
\newblock \bibinfo{journal}{\bibinfo{title}{Learning a deep single image contrast enhancer from multi-exposure images}}.
\newblock {\emph{\JournalTitle{IEEE Transactions on Image Processing}}} \textbf{\bibinfo{volume}{27}}, \bibinfo{pages}{2049--2062} (\bibinfo{year}{2018}).

\bibitem{yu2020bdd100k}
\bibinfo{author}{Yu, F.} \emph{et~al.}
\newblock \bibinfo{journal}{\bibinfo{title}{Bdd100k: A diverse driving video database with scalable annotation tooling}}.
\newblock {\emph{\JournalTitle{arXiv preprint arXiv:1805.04687}}}  (\bibinfo{year}{2020}).

\bibitem{cordts2016cityscapes}
\bibinfo{author}{Cordts, M.} \emph{et~al.}
\newblock \bibinfo{title}{The cityscapes dataset for semantic urban scene understanding}.
\newblock In \emph{\bibinfo{booktitle}{Proceedings of the IEEE conference on computer vision and pattern recognition}}, \bibinfo{pages}{3213--3223} (\bibinfo{year}{2016}).

\bibitem{lee2012contrast}
\bibinfo{author}{Lee, C.} \& \bibinfo{author}{Kim, C.-S.}
\newblock \bibinfo{journal}{\bibinfo{title}{Contrast enhancement based on layered difference representation of 2d histograms}}.
\newblock {\emph{\JournalTitle{IEEE Transactions on Image Processing}}} \textbf{\bibinfo{volume}{22}}, \bibinfo{pages}{5372--5384} (\bibinfo{year}{2013}).

\bibitem{guo2016lime}
\bibinfo{author}{Guo, X.}, \bibinfo{author}{Li, Y.} \& \bibinfo{author}{Ling, H.}
\newblock \bibinfo{journal}{\bibinfo{title}{Lime: Low-light image enhancement via illumination map estimation}}.
\newblock {\emph{\JournalTitle{arXiv preprint arXiv:1605.09782}}}  (\bibinfo{year}{2016}).

\bibitem{wei2018deep}
\bibinfo{author}{Wei, C.}, \bibinfo{author}{Wang, W.}, \bibinfo{author}{Yang, W.} \& \bibinfo{author}{Liu, J.}
\newblock \bibinfo{title}{Deep retinex decomposition for low-light enhancement}.
\newblock In \emph{\bibinfo{booktitle}{British Machine Vision Conference (BMVC)}} (\bibinfo{year}{2018}).

\bibitem{yang2021fidelity}
\bibinfo{author}{Yang, W.}, \bibinfo{author}{Wang, S.}, \bibinfo{author}{Fang, Y.}, \bibinfo{author}{Wang, Y.} \& \bibinfo{author}{Liu, J.}
\newblock \bibinfo{title}{From fidelity to perceptual quality: A semi-supervised approach for low-light image enhancement}.
\newblock In \emph{\bibinfo{booktitle}{Proceedings of the IEEE/CVF Conference on Computer Vision and Pattern Recognition}}, \bibinfo{pages}{3066--3075} (\bibinfo{year}{2021}).

\bibitem{ma2015perceptual}
\bibinfo{author}{Ma, K.}, \bibinfo{author}{Zeng, K.} \& \bibinfo{author}{Wang, Z.}
\newblock \bibinfo{journal}{\bibinfo{title}{Perceptual quality assessment for multi-exposure image fusion}}.
\newblock {\emph{\JournalTitle{IEEE Transactions on Image Processing}}} \textbf{\bibinfo{volume}{24}}, \bibinfo{pages}{3345--3356} (\bibinfo{year}{2015}).

\bibitem{wang2013naturalness}
\bibinfo{author}{Wang, Y.}, \bibinfo{author}{Wang, Q.} \& \bibinfo{author}{Liao, Q.}
\newblock \bibinfo{journal}{\bibinfo{title}{Naturalness preserved enhancement algorithm for non-uniform illumination images}}.
\newblock {\emph{\JournalTitle{IEEE Transactions on Image Processing}}} \textbf{\bibinfo{volume}{22}}, \bibinfo{pages}{3538--3548} (\bibinfo{year}{2013}).

\bibitem{VV}
\bibinfo{author}{Jinda-Apiraksa, A.}, \bibinfo{author}{Vonikakis, V.} \& \bibinfo{author}{Winkler, S.}
\newblock \bibinfo{title}{California-nd: An annotated dataset for near-duplicate detection in personal photo collections}.
\newblock In \emph{\bibinfo{booktitle}{2013 Fifth International Workshop on Quality of Multimedia Experience (QoMEX)}}, \bibinfo{pages}{142--147} (\bibinfo{organization}{IEEE}, \bibinfo{year}{2013}).

\bibitem{wang2004image}
\bibinfo{author}{Wang, Z.}, \bibinfo{author}{Bovik, A.~C.}, \bibinfo{author}{Sheikh, H.~R.} \& \bibinfo{author}{Simoncelli, E.~P.}
\newblock \bibinfo{journal}{\bibinfo{title}{Image quality assessment: from error visibility to structural similarity}}.
\newblock {\emph{\JournalTitle{IEEE Transactions on Image Processing}}} \textbf{\bibinfo{volume}{13}}, \bibinfo{pages}{600--612} (\bibinfo{year}{2004}).

\bibitem{hore2010image}
\bibinfo{author}{Hore, A.} \& \bibinfo{author}{Ziou, D.}
\newblock \bibinfo{title}{Image quality metrics: Psnr vs. ssim}.
\newblock In \emph{\bibinfo{booktitle}{2010 20th International Conference on Pattern Recognition}}, \bibinfo{pages}{2366--2369} (\bibinfo{organization}{IEEE}, \bibinfo{year}{2010}).

\bibitem{zhang2011fsim}
\bibinfo{author}{Zhang, L.}, \bibinfo{author}{Zhang, L.}, \bibinfo{author}{Mou, X.} \& \bibinfo{author}{Zhang, D.}
\newblock \bibinfo{journal}{\bibinfo{title}{Fsim: A feature similarity index for image quality assessment}}.
\newblock {\emph{\JournalTitle{IEEE Transactions on Image Processing}}} \textbf{\bibinfo{volume}{20}}, \bibinfo{pages}{2378--2386} (\bibinfo{year}{2011}).

\bibitem{zhang2014vsi}
\bibinfo{author}{Zhang, L.}, \bibinfo{author}{Shen, Y.} \& \bibinfo{author}{Li, H.}
\newblock \bibinfo{journal}{\bibinfo{title}{Vsi: A visual saliency-induced index for perceptual image quality assessment}}.
\newblock {\emph{\JournalTitle{IEEE Transactions on Image Processing}}} \textbf{\bibinfo{volume}{23}}, \bibinfo{pages}{4270--4281} (\bibinfo{year}{2014}).

\bibitem{zhang2018unreasonable}
\bibinfo{author}{Zhang, R.}, \bibinfo{author}{Isola, P.}, \bibinfo{author}{Efros, A.~A.}, \bibinfo{author}{Shechtman, E.} \& \bibinfo{author}{Wang, O.}
\newblock \bibinfo{journal}{\bibinfo{title}{The unreasonable effectiveness of deep features as a perceptual metric}}.
\newblock {\emph{\JournalTitle{Proceedings of the IEEE Conference on Computer Vision and Pattern Recognition (CVPR)}}} \bibinfo{pages}{586--595} (\bibinfo{year}{2018}).

\bibitem{ding2020image}
\bibinfo{author}{Ding, K.}, \bibinfo{author}{Ma, K.}, \bibinfo{author}{Wang, S.} \& \bibinfo{author}{Simoncelli, E.~P.}
\newblock \bibinfo{journal}{\bibinfo{title}{Image quality assessment: Unifying structure and texture similarity}}.
\newblock {\emph{\JournalTitle{IEEE Transactions on Pattern Analysis and Machine Intelligence}}}  (\bibinfo{year}{2020}).
\newblock \bibinfo{note}{Early access}.

\bibitem{NIMA}
\bibinfo{author}{Talebi, H.} \& \bibinfo{author}{Milanfar, P.}
\newblock \bibinfo{title}{{NIMA}: Neural image assessment}.
\newblock In \emph{\bibinfo{booktitle}{Proceedings of the IEEE Conference on Computer Vision and Pattern Recognition (CVPR)}}, \bibinfo{pages}{5168--5177} (\bibinfo{year}{2018}).

\bibitem{PaQ2PiQ}
\bibinfo{author}{Ying, X.} \emph{et~al.}
\newblock \bibinfo{title}{{PaQ-2-PiQ}: Attribute-aware learning for blind image quality assessment}.
\newblock In \emph{\bibinfo{booktitle}{Proceedings of the IEEE/CVF Conference on Computer Vision and Pattern Recognition (CVPR)}}, \bibinfo{pages}{3560--3569} (\bibinfo{year}{2020}).

\bibitem{DBCNN}
\bibinfo{author}{Zhang, L.}, \bibinfo{author}{Li, H.}, \bibinfo{author}{Fu, X.}, \bibinfo{author}{Xiong, S.} \& \bibinfo{author}{Dong, W.}
\newblock \bibinfo{title}{{DBCNN}: A dual branch convolutional neural network for no-reference image quality assessment}.
\newblock In \emph{\bibinfo{booktitle}{Proceedings of the IEEE International Conference on Computer Vision (ICCV)}}, \bibinfo{pages}{5660--5669}, \doiprefix\url{10.1109/ICCV.2018.00593} (\bibinfo{year}{2018}).

\bibitem{Yang2022MANIQA}
\bibinfo{author}{Yang, H.}, \bibinfo{author}{Zhu, P.}, \bibinfo{author}{Wang, Z.}, \bibinfo{author}{Min, X.} \& \bibinfo{author}{Mou, X.}
\newblock \bibinfo{title}{{MANIQA}: Multi-dimension attention network for no-reference image quality assessment}.
\newblock In \emph{\bibinfo{booktitle}{Proceedings of the IEEE/CVF Conference on Computer Vision and Pattern Recognition (CVPR)}}, \bibinfo{pages}{11432--11441}, \doiprefix\url{10.1109/CVPR52688.2022.01115} (\bibinfo{year}{2022}).

\bibitem{Wang2023CLIPIQA}
\bibinfo{author}{Wang, Z.}, \bibinfo{author}{Lin, R.}, \bibinfo{author}{Lu, X.} \& \bibinfo{author}{Wang, Z.}
\newblock \bibinfo{title}{{CLIP-IQA}: Clip-based image quality assessment}.
\newblock In \emph{\bibinfo{booktitle}{Proceedings of the IEEE/CVF Conference on Computer Vision and Pattern Recognition (CVPR)}}, \bibinfo{pages}{3565--3574}, \doiprefix\url{10.1109/CVPR52729.2023.00328} (\bibinfo{year}{2023}).

\bibitem{Lin2022TReS}
\bibinfo{author}{Lin, S.}, \bibinfo{author}{Wang, Q.}, \bibinfo{author}{Jiang, J.} \& \bibinfo{author}{Ma, J.}
\newblock \bibinfo{title}{Tres: A transformer relation network for no-reference image quality assessment}.
\newblock In \emph{\bibinfo{booktitle}{European Conference on Computer Vision (ECCV)}}, vol. \bibinfo{volume}{13671} of \emph{\bibinfo{series}{Lecture Notes in Computer Science}}, \bibinfo{pages}{267--284}, \doiprefix\url{10.1007/978-3-031-19800-2\_16} (\bibinfo{publisher}{Springer}, \bibinfo{year}{2022}).
\newblock \bibinfo{note}{We use the {KONIQ}-fine-tuned model}.

\bibitem{Su2020HyperIQA}
\bibinfo{author}{Su, M.~Y.}, \bibinfo{author}{Zeng, D.}, \bibinfo{author}{Hong, Z.}, \bibinfo{author}{Ouyang, W.} \& \bibinfo{author}{Yu, X.}
\newblock \bibinfo{journal}{\bibinfo{title}{Blindly assess image quality in the wild leveraging an uncertainty-aware {HyperNet}}}.
\newblock {\emph{\JournalTitle{IEEE Transactions on Image Processing}}} \textbf{\bibinfo{volume}{29}}, \bibinfo{pages}{5035--5048}, \doiprefix\url{10.1109/TIP.2020.2985256} (\bibinfo{year}{2020}).

\bibitem{khalid2021gaussian}
\bibinfo{author}{Khalid, H.}, \bibinfo{author}{Ali, M.} \& \bibinfo{author}{Ahmed, N.}
\newblock \bibinfo{journal}{\bibinfo{title}{Gaussian process-based feature-enriched blind image quality assessment}}.
\newblock {\emph{\JournalTitle{Journal of Visual Communication and Image Representation}}} \textbf{\bibinfo{volume}{77}}, \bibinfo{pages}{103092} (\bibinfo{year}{2021}).

\bibitem{aslam2024qualitynet}
\bibinfo{author}{Aslam, M.~A.} \emph{et~al.}
\newblock \bibinfo{journal}{\bibinfo{title}{Qualitynet: A multi-stream fusion framework with spatial and channel attention for blind image quality assessment}}.
\newblock {\emph{\JournalTitle{Scientific Reports}}} \textbf{\bibinfo{volume}{14}}, \bibinfo{pages}{26039} (\bibinfo{year}{2024}).

\bibitem{ahmed2021piqi}
\bibinfo{author}{Ahmed, N.}, \bibinfo{author}{Asif, H. M.~S.} \& \bibinfo{author}{Khalid, H.}
\newblock \bibinfo{journal}{\bibinfo{title}{Piqi: perceptual image quality index based on ensemble of gaussian process regression}}.
\newblock {\emph{\JournalTitle{Multimedia Tools and Applications}}} \textbf{\bibinfo{volume}{80}}, \bibinfo{pages}{15677--15700} (\bibinfo{year}{2021}).

\bibitem{mittal2012no}
\bibinfo{author}{Mittal, A.}, \bibinfo{author}{Moorthy, A.~K.} \& \bibinfo{author}{Bovik, A.~C.}
\newblock \bibinfo{journal}{\bibinfo{title}{No-reference image quality assessment in the spatial domain}}.
\newblock {\emph{\JournalTitle{IEEE Transactions on Image Processing}}} \textbf{\bibinfo{volume}{21}}, \bibinfo{pages}{4695--4708} (\bibinfo{year}{2012}).

\bibitem{zhang2021blind}
\bibinfo{author}{Zhang, W.}, \bibinfo{author}{Chen, C.}, \bibinfo{author}{Li, C.} \& \bibinfo{author}{Ma, K.}
\newblock \bibinfo{title}{Blind image quality assessment using a score distribution prior}.
\newblock In \emph{\bibinfo{booktitle}{Proceedings of the IEEE/CVF Conference on Computer Vision and Pattern Recognition (CVPR)}}, \bibinfo{pages}{16758--16767} (\bibinfo{year}{2021}).

\bibitem{ahmed2020perceptual}
\bibinfo{author}{Ahmed, N.} \& \bibinfo{author}{Asif, H. M.~S.}
\newblock \bibinfo{journal}{\bibinfo{title}{Perceptual quality assessment of digital images using deep features}}.
\newblock {\emph{\JournalTitle{Computing and Informatics}}} \textbf{\bibinfo{volume}{39}}, \bibinfo{pages}{385--409} (\bibinfo{year}{2020}).

\bibitem{BIMEF}
\bibinfo{author}{Ying, Z.}, \bibinfo{author}{Li, G.} \& \bibinfo{author}{Gao, W.}
\newblock \bibinfo{journal}{\bibinfo{title}{A bio-inspired multi-exposure fusion framework for low-light image enhancement}}.
\newblock {\emph{\JournalTitle{arXiv preprint arXiv:1711.00591}}}  (\bibinfo{year}{2017}).

\bibitem{LIME}
\bibinfo{author}{Guo, X.}, \bibinfo{author}{Li, Y.} \& \bibinfo{author}{Ling, H.}
\newblock \bibinfo{journal}{\bibinfo{title}{Lime: Low-light image enhancement via illumination map estimation}}.
\newblock {\emph{\JournalTitle{IEEE Transactions on image processing}}} \textbf{\bibinfo{volume}{26}}, \bibinfo{pages}{982--993} (\bibinfo{year}{2016}).

\bibitem{MF}
\bibinfo{author}{Fu, X.} \emph{et~al.}
\newblock \bibinfo{journal}{\bibinfo{title}{A fusion-based enhancing method for weakly illuminated images}}.
\newblock {\emph{\JournalTitle{Signal Processing}}} \textbf{\bibinfo{volume}{129}}, \bibinfo{pages}{82--96} (\bibinfo{year}{2016}).

\bibitem{MultiscaleRetinex}
\bibinfo{author}{Petro, A.~B.}, \bibinfo{author}{Sbert, C.} \& \bibinfo{author}{Morel, J.-M.}
\newblock \bibinfo{journal}{\bibinfo{title}{Multiscale retinex}}.
\newblock {\emph{\JournalTitle{Image processing on line}}} \bibinfo{pages}{71--88} (\bibinfo{year}{2014}).

\bibitem{jiang2021enlightengan}
\bibinfo{author}{Jiang, Y.} \emph{et~al.}
\newblock \bibinfo{journal}{\bibinfo{title}{Enlightengan: Deep light enhancement without paired supervision}}.
\newblock {\emph{\JournalTitle{IEEE Transactions on Image Processing}}} \textbf{\bibinfo{volume}{30}}, \bibinfo{pages}{2340--2349} (\bibinfo{year}{2021}).

\end{thebibliography}
\end{document}